\documentclass[11pt]{article}

\usepackage{amsmath}
\usepackage{graphicx,psfrag,epsf}
\usepackage{enumerate}
\usepackage[numbers, sort&compress]{natbib}
\usepackage[hyphens]{url} %
\usepackage{ulem} %
\usepackage{standalone}

\newcommand{\blind}{0}

\usepackage{makecell}
\usepackage[table,dvipsnames]{xcolor}
\usepackage{amsthm}
\usepackage{bm}
\usepackage{bbm}
\usepackage{mathtools}
\usepackage{collcell}
\usepackage{dsfont}
\usepackage{rotating}
\usepackage{wrapfig}
\usepackage[utf8]{inputenc}
\usepackage{tikz}
\usepackage{lettrine}
\usepackage{dblfloatfix}
\usepackage[english]{babel} 
\usepackage{amsfonts}
\graphicspath{ {fig/} }
\usepackage[hyperfootnotes=false]{hyperref}

\usepackage{subfigure}
\usepackage{arydshln}
\usepackage{lipsum}
\usetikzlibrary{arrows,decorations.markings}
\usepackage{standalone}

\newtheorem{theorem}{Theorem}

\definecolor{gray}{rgb}{0.5,0.5,0.5}
\definecolor{red}{rgb}{0.8,0,0}
\definecolor{dred}{rgb}{0.5,0,0}
\definecolor{blue}{rgb}{0,0.1,1}
\definecolor{dblue}{rgb}{0,0.1,0.6}
\definecolor{cyan}{rgb}{0,0.5,.5}
\definecolor{dcyan}{rgb}{0,0.3,.3}
\definecolor{mpurple}{rgb}{.7,0,.9}

\definecolor{b}{rgb}{0,0,.8}	%
\definecolor{g}{rgb}{0,.6,0}	%
\definecolor{n}{rgb}{0,0,0}	%
\definecolor{h}{rgb}{0.4,0.2,0.2}	%
\definecolor{v}{rgb}{0.2,0.6,0}

\newcommand{\bsa}{\boldsymbol a}
\newcommand{\bsb}{\boldsymbol b}

\newcommand{\bse}{\boldsymbol e}

\newcommand{\bsp}{\boldsymbol p}
\newcommand{\bsq}{\boldsymbol q}

\newcommand{\bsw}{\boldsymbol w}

\newcommand{\bsy}{\boldsymbol y}

\newcommand{\bsA}{\boldsymbol A}
\newcommand{\bsB}{\boldsymbol B}

\newcommand{\bsD}{\boldsymbol D}
\newcommand{\bsE}{\boldsymbol E}

\newcommand{\bsI}{\boldsymbol I}

\newcommand{\bsL}{\boldsymbol L}

\newcommand{\bsP}{\boldsymbol P}
\newcommand{\bsQ}{\boldsymbol Q}

\newcommand{\bsS}{\boldsymbol S}

\newcommand{\bsU}{\boldsymbol U}

\newcommand{\bsW}{\boldsymbol W}

\newcommand{\bsone}{\boldsymbol 1}

\newcommand{\bsnull}{\boldsymbol 0}

\newcommand{\bsNull}{\text{\textbf{O}}}

\newcommand{\bsbeta}{\boldsymbol \beta}

\newcommand{\bsPhi}{\boldsymbol \Phi}

\DeclareMathOperator*{\argmin}{arg\,min}

\newcommand{\ov}\overline
\newcommand{\what}{\widehat}
\newcommand{\wtilde}{\widetilde}

\newcommand{\rig}\right
\newcommand{\lef}\left
\newcommand{\nf}\normalfont

\usepackage{forest} %
\usepackage{multirow}

 \newcommand{\FZ}[1]{\color{black} #1 \color{black}} %

 \newcommand{\PG}[1]{\color{black} #1 \color{black}} %

  \newcommand{\PGG}[1]{\color{black} #1 \color{black}} %

\setcitestyle{authoryear,comma,aysep{,}}

\begin{document}

	\def\spacingset#1{\renewcommand{\baselinestretch}%
		{#1}\small\normalsize} \spacingset{1}
	
	\if0\blind
	{
		\title{\bf Hierarchical forecasting for aggregated curves with an application to day-ahead electricity price auctions}
		\author{Paul Ghelasi\footnote{Corresponding author. Email addresses: paul.ghelasi@mail.com; paul.ghelasi@stud.uni-due.de} , Florian Ziel\footnote{Chair of Environmental Economics, esp. Economics of Renewable Energy. Email address: florian.ziel@uni-due.de}\\
			University of Duisburg-Essen, Germany}
		\maketitle
	} \fi
	
	\if1\blind
	{
		
		\bigskip
		
		\begin{center}
			{\LARGE\bf  Hierarchical forecasting for aggregated curves with an application to day-ahead electricity price auctions}
		\end{center}
		
	} \fi

	\begin{abstract}
		Aggregated curves are common structures in economics and finance, and the most prominent examples are supply and demand curves. In this study, we \PG{exploit the fact} that all aggregated curves have an intrinsic hierarchical structure, and thus hierarchical reconciliation methods can be used to improve the forecast accuracy. \FZ{We provide an in-depth theory on how aggregated curves can be constructed or deconstructed, and conclude that these methods are equivalent under weak assumptions.}
		We consider multiple reconciliation methods for aggregated curves, including previously established bottom-up, top-down, and linear optimal reconciliation approaches. \PG{We also present a new benchmark reconciliation method called 'aggregated-down' with similar complexity to bottom-up and top-down approaches, but it tends to provide better accuracy in this setup.}
		We conducted an empirical forecasting study on the German day-ahead power auction market by predicting the demand and supply curves, where their equilibrium determines the electricity price for the next day. Our results demonstrate that hierarchical reconciliation methods can be used to improve the forecasting accuracy of aggregated curves.
	\end{abstract}
	
	\noindent%
	\vfill
	
	\spacingset{1.45} %

	\section{Introduction and motivation}
	
	Curves are frequently encountered structures in \FZ{various scientific disciplines, and especially in economics and finance. Prominent examples are} supply and demand curves \citep{marshall2009principles, mankiw2014principles, pindyck1995microeconomics}. Other well-known examples are forwards and futures price curves \citep{hull2003options}, yield curves \citep{gurkaynak2007us}, Engel curves \citep{aitchison1954synthesis, banks1997quadratic}, Philips curve \citep{phillips}, and various types of cost curves \citep{eiteman1952shape, pindyck1995microeconomics}. \PG{The accurate estimation of these curves is important because other measures such as equilibrium values can be derived from them, and modeling a curve as a whole has the benefit of preserving additional information such as its shape or slope, which could be very useful in deriving the strategies of market participants, equilibrium values, or probabilistic forecasts, rather than modeling the equilibrium prices directly \citep{xmodel}.}
	
\begin{figure}[!htpb]
		\centering
		\includegraphics[width=0.6\linewidth]{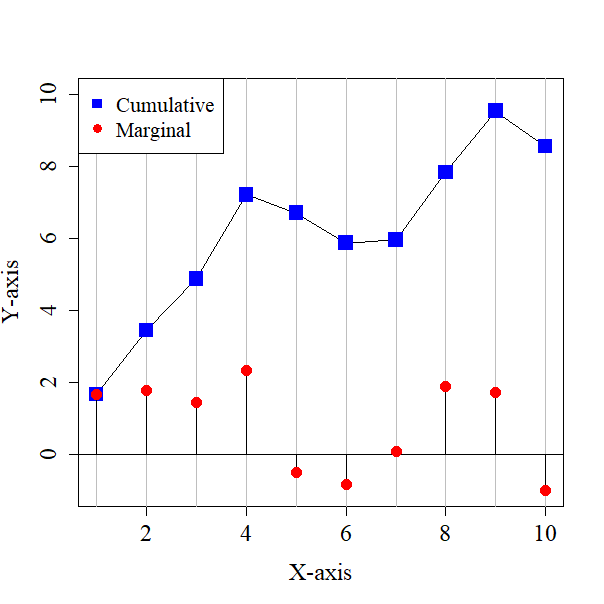}
		\caption{Example curve with marginal steps}
		\scriptsize{}
		\label{fig:simcurve}
	\end{figure}

	\PG{It is important to note that we refer to curves as a single observation at a point in time, as is the case in functional data analysis (FDA) \citep{shang2017grouped, hyndman2007robust, shang2011nonparametric}.} \FZ{In this study, we consider finite-dimensional representations of curves given by a finite grid of $x$-values and their corresponding $y$-values, often referred to as $f(x)$.
	} Every curve with well-defined increments can be readily aggregated or disaggregated into marginal and cumulative values. For example, Figure \ref{fig:simcurve} shows a simulated curve and its corresponding marginal values at each step. The cumulative values can be obtained from the marginal values by cumulatively summing them up, and the marginal values can be obtained from the cumulative values by \FZ{discrete} differentiation. \FZ{Hence, from a computational point of view, a curve is simply a vector of aggregated values. }	
	This inherent structure of aggregated curves can be interpreted as a hierarchy, and thus hierarchical time series theory can be applied to potentially improve the forecasting accuracy \citep{hyndman2018forecasting}. 
	
	An hierarchical time series is a set of time series that naturally relate to each other, i.e. sum up or break down, according to a certain logic \PG{\citep{wickramasuriya2019optimal, hyndman2018forecasting, spiliotis2021hierarchical}. This property is referred to as \textit{coherency} and it is generally not given for individual forecasts. Approaches that allow independently forecasted time series at certain or all levels of the hierarchy to be made coherent are referred to as \textit{reconciliation} methods. 
	}

	\FZ{
	Notable advances have been recently made in optimal reconciliation approaches as well as in improving classic approaches, such as top-down and bottom-up approaches, in terms of their forecasting accuracy \cite{hyndman2018forecasting}. Most notable is the optimal \textit{minimum trace} reconciling method \cite{wickramasuriya2019optimal}.
	Other methods have also been proposed, such as a \textit{game-theoretically optimal} reconciliation approach \citep{van2015game}, averaging approaches called \textit{level conditional coherent} (LCC) and \textit{combined conditional coherent} (CCC) point forecasts \citep{di2021forecast,hollyman2021understanding} and machine-learning based reconciliation \citep{spiliotis2021hierarchical}, \citep{bregere2022online}, \citep{huard2020hierarchical}.}

	\PG{In this study, we introduce four novel features into the field of hierarchical forecasting. %
	First, we exploit the fact that all aggregated curves have an implicit hierarchical structure and that reconciliation approaches can be used to potentially improve their forecast accuracy. To our knowledge, this is the first time that aggregated curves have been treated as hierarchical structures. We consider various methods of constructing and deconstruction the curves, including different representations of aggregated curves. %
	Second, we introduce a new, simple reconciliation approach called \textit{aggregated-down} with similar complexity to the top-down approach, which we recommend using as a benchmark method alongside bottom-up and top-down approaches. Third, we study minimum trace optimal reconciliation approaches for aggregated curves in detail. In particular, we provide a result to show that under some assumptions, the reconciliation approach is independent of the representation of the curve. Finally, we applied all of these approaches in an empirical setting to forecast the supply and demand curves for day-ahead electricity price auctions. We conclude that forecast accuracy can be improved through hierarchical reconciliation.}

	\PG{To assess the effect of hierarchical reconciliation approaches on aggregated curves in an empirical setting we consider an application to the German day-ahead electricity markets. 
	In this auction-based market, submitted bids are aggregated to form supply and demand curves \citep{narajewski2022optimal}.
	The intersection of the curves yields the day-ahead electricity price. Hence, the market mechanism itself fits our idea of aggregated curves, and thus the application of hierarchical forecasting is a natural extension. }
	
\FZ{	\cite{xmodel, ziel2018probabilistic, haben2021probabilistic} forecast the day-ahead electricity price by forecasting the corresponding bids, aggregating them to form supply and demand curves, and computing the resulting intersection of these curves to yield the final price forecasts. This implicit application of the simple bottom-up reconciliation approach improved both point and probabilistic price forecasts.}
	Recent related papers \cite{kulakov2020x, mestre2020forecasting, 
		soloviova2021efficient} %
	which forecast electricity supply and demand curves show similar findings.
	
	The remainder of this paper is organized as follows. In \hyperref[sec:hierarchy]{Section 2}, we introduce the notation for the inherent hierarchical structure of aggregated curves. \PG{We also describe different ways to construct and represent curves via aggregation and/or disaggregation. In \hyperref[sec:reconciliation]{Section 3}, we present the reconciliation approaches that we consider in this study, including the new aggregated-down approach and their simplified formulas according to the specific hierarchical structure of aggregated curves. Furthermore, we show that under some assumptions, the reconciliation approach is independent of the previously introduced curve representations. In 
	\hyperref[sec:simulation]{Section 4}, we present a simulation study of the new aggregated-down approach where we analyze and compare with the other reconciliation methods considered in this study.}
	In \hyperref[sec:Application]{Section 5}, we introduce the market clearing mechanism for the German day-ahead electricity market and the model used to forecast the day-ahead supply and demand curves. We also present our empirical study, the data used, and the empirical results. We summarize the results and provide our conclusions in \hyperref[sec:conclusions]{Section 6}.

\section{Hierarchical structure of aggregated curves}
\label{sec:hierarchy}

\FZ{The hierarchical structure of a curve can be represented in multiple ways, i.e. the relationship between the bottom-level or marginal values and the curve itself, or the aggregated values. The curve can be obtained from an aggregation procedure with the marginals. Alternatively, we can start with the curve and describe a disaggregation relationship to obtain the bottom values.
	
However, a natural method of representation that we refer to as the \emph{canonical structure} of an aggregated curve is shown in Figure \ref{fig:hierarch}. At the end of this section, we also discuss other representations.
}

\subsection{Canonical representation}
\begin{figure}[!htpb]
	\centering
	\scalebox{0.65}{\LARGE
		\begin{forest}
			for tree={
				draw,
				circle,
				if n children=0{tier=terminal}{},
				s sep=40pt,
				l sep+=-15pt,
			},
			before typesetting nodes={
				for tree={
					shape=circle,
					minimum size = 50,
				}
			}
			[$a_{n}$, blue,
			[$a_{n-1}$, blue [$\dots$,
			[$a_{2}$, blue,
			[$\overset{{\color{blue}a_1}=}{{\color{red} b_{1}}}$, mpurple]
			[$b_{2}$, red]
			]
			[2.5, white, edge = white]
			]
			[$b_{n-1}$, red]
			]
			[$b_n$, red]]
	\end{forest}}
	\caption{Hierarchical structure of an aggregated curve}
	\label{fig:hierarch}
\end{figure}
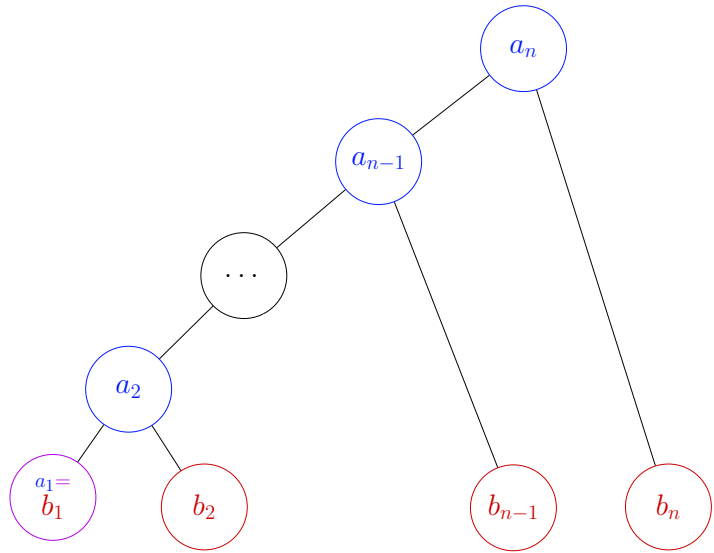

\FZ{
The canonical representation of an aggregated curve at we illustrate in Figure \ref{fig:hierarch} is regarded as the standard representation. This representation also appears naturally when modeling auction curves in economics \citep{xmodel}. It can be obtained by considering the bottom values together with an aggregation procedure to receive the aggregated values, but also the other way around.
}

\FZ{
Based on the first approach above, we introduce the following notation. Let $\bsb = ( b_{1} , \dots , b_{n} )'$ be the $n$-dimensional vector of bottom-level or marginal values with $n>1$, as shown in Figure \ref{fig:hierarch}. Now, we introduce their aggregation by the vector $\bsa = ( a_{1} , \dots , a_{n} )'$, which is an $n$-dimensional vector of aggregated or cumulative values starting from the top level, where $a_i$ for $i \in \{1,\dots,n\}$ represents the curve at point $i$. Formally, this is defined as
\begin{equation}
a_i = \sum^{i}_{j=1} b_j \textrm ,
\label{eq:curvecum_b_to_a}
\end{equation}
i.e. the cumulative sum of $\bsb$.
In addition, the recursive relationship holds
\begin{equation}
a_{i} = a_{i-1} + b_{i} \text{ for } 1 \le i<n	\text{ and }
a_{1} = b_1.
\label{eq:curvecum_b_to_a_recursive}
\end{equation}
We note that Equation (\ref{eq:curvecum_b_to_a_recursive}) is simply the mathematical representation of Figure \ref{fig:hierarch}. The top level $a_{n}$ is the most aggregated level. 

Alternatively, we can introduce the canonical representation starting from the aggregated values $\bsa$. Then, we can obtain the bottom values $\bsb$ by disaggregation or differencing 
with an initial value for $b_1$: %
\begin{equation}
b_i =  a_{i}-a_{i-1} \textrm ,
\text{ for } 1 < i\leq n	\text{ and }
b_{1} = a_1.
\label{eq:curvecum_a_to_b}
\end{equation}
Furthermore, we can express $\bsb$ as $\bsb = \bsD \bsa$ where $\bsD$ is an invertible $n$-dimensional quadratic matrix defined as
\begin{equation}
	\bsD_n = \begin{bmatrix}
	1 & 0 & \ldots &0&0 \\
	-1 & 1 & \ldots &0&0 \\
	\vdots  & \vdots & \ddots &\vdots &\vdots  \\
	0 & 0 & \ldots &-1&1 
\end{bmatrix}.
\label{eq_a_to_b}
\end{equation}
Consequently, the matrix representation of equation \eqref{eq:curvecum_b_to_a} is $\bsa = \bsD_n^{-1}\bsb$ where $\bsD_n^{-1}$ is a lower-left triangular matrix with ones on the diagonal and lower triangle.
}
s
\FZ{
Using the definitions of aggregated values $\bsa$ and bottom-level values $\bsb$ we can compactly write all of the values shown in Figure \eqref{fig:hierarch} as the $(2n-1)$-dimensional vector 
 \begin{equation} \bsy = 
 \begin{bmatrix} a_n \\ \vdots \\a_2 \\ \bsb \end{bmatrix}
 \label{eq_def_y}
 \end{equation}
 which contains the values of $\bsa$ in inverse order except for its first value $a_1$, which matches $b_1$.
Thus, $(y_1,\ldots, y_n)'$ may be regarded as aggregated values and $(y_n,\ldots, y_{2n-1})'$ as bottom-level values. 
}

The aggregation relationship within $\bsy$ can be represented using the $(2n-1)\times n$-dimensional \emph{summation matrix} $\bsS$, which describes the hierarchical structure of the considered data such that $\bsy = \bsS \bsb$ holds.
\FZ{For the considered canonical representation we have
 \begin{equation}
	\bsS = \left[\begin{array}{c:c} 
		\bsone_{n-1} & \bsU_{n-1} \\ \hdashline[2pt/2pt] 
		\multicolumn{2}{c}{\bsI_n}  \end{array}\right]
= \left[ \begin{array} {ccccc}
	\multicolumn{1}{c:}{1} & 1 &\cdots & 1 & 1 \\
	\multicolumn{1}{c:}{1} & 1 &\cdots & 1 & 0 \\
	\multicolumn{1}{c:}{\vdots} & \vdots &\ddots & \vdots & \vdots \\
	\multicolumn{1}{c:}{1}  & 1 &\cdots & 0 & 0 \\ \hdashline[2pt/2pt] 
1 & 0 &\cdots & 0 & 0 \\ 
0 & 1 &\cdots & 0 & 0 \\
\vdots & \vdots &\ddots & \vdots & \vdots \\
0 & 0 &\cdots & 1 & 0 \\
0 & 0 &\cdots & 0 & 1 \\
 \end{array}\right]
 ,
 \label{eq_summation_matrix_S}
\end{equation}
where $\bsU_{n-1}$ is a unit anti-diagonal upper-left triangular matrix of dimension $(n-1)$ and $\bsI_n$ is an identity matrix of dimension $n$.
}

\FZ{Other situations can be considered where even more parts of the hierarchical curve are available. Using the notation $b_{i:j} = \sum_{l=i}^j b_l$ the canonical setting uses $b_{1:j}$ for $j>1$ for the aggregated values, i.e. $\bsa = (b_{1:1}, b_{1:2},\ldots, b_{1:n})'$. We could also consider a setting where, e.g. $b_{2:3} = b_2+b_3$, is also available, which would enrich the hierarchical structure with this additional information. However, the resulting structures that include general $b_{i:j}$ combinations could include up to $n(n-1)/2$ values in the corresponding $\bsy$ vector. This would increase computational costs substantially, so we only study the specific structure where forecasts $\what{\bsy}$ (or equivalently $\what{\bsa}$ and $\what{\bsb}$) are available to the forecaster.}

\subsection{Other representations of aggregated curves}

\FZ{In this subsection, we discuss representations other than the canonical one for aggregated curves. We generalize the representation scheme so the canonical representation is embedded. In particular, we consider a situation where the aggregated values $\bsa$ are given and the scope involves defining a disaggregation rule to receive the bottom values $\bsb$. 

We introduce $k$ as the element of $\bsa$ where from we start the disaggregation rule.	
In the canonical representation we disaggregate $\bsa$ starting from the first value $k=1$ of the curve, i.e. for $k=1$ it holds that $b_k=a_k$, $b_i= a_{i} - a_{i-1}$ for $i>k$ (see Equation \eqref{eq:curvecum_a_to_b}).
This starting point of aggregation/disaggregation $k$ could be changed to achieve an alternative bottom-values vector $\bsb_{[k]}$. Clearly, $\bsb = \bsb_{[1]}$ and if we start disaggregating from the end, i.e. $k=n$, we obtain $b_{[n],1}= a_n$, $b_{[n],2}= a_{n-1}-a_{n}$, and in general, $b_{[n],i}=a_{n-i+1} - a_{n-i+2}$. Thus, $b_{[n],i}$ has the opposite sign to $b_i$ for $i<n$.
However, this approach can be embedded in the canonical representation, which we receive by defining $b_i = b_{[n],n-i+1}$. Thus, the theory of canonical representations can also be applied.

If we consider the disaggregation procedure with an initial value at $k$ where $1<k<n$ with the corresponding value $b_{[k],k} = a_{k}$, 
 then we obtain a representation 
 that is substantially different from the $\bsb_{[1]}$ and $\bsb_{[n]}$ situations. 
The reason for this difference is that we require two directions of aggregation, with one for bottom values larger than $k$ and the other one for smaller values. 
Thus, we have
\begin{equation}
	b_{[k],i}= \begin{cases}
		a_{i} - a_{i-1}	& \text{, if } i>k \\
		a_{i} - a_{i+1}	& \text{, if } i<k \\
		a_{i}	& \text{, if } i=k
	\end{cases}
\label{eq_def_bki}
\end{equation}
 We observe that the special cases of $\bsb_{[1]}$ (the canonical representation) and $\bsb_{[n]}$ can be defined using the definition \eqref{eq_def_bki}.
 For illustrative purposes and to easier understand, we provide an
$n=6$-dimensional example for $k=1,3,6$ in Table \ref{tab_hierarch_example}.
 \begin{table}
	\begin{center}		
 \begin{tabular}{r|r|cccccc}
	&$i$ &	1 & 2 & 3 & 4 & 5 & 6 \\ \hline
	$\bsa$& $a_i$ &	1 & 4 & 6 & 7 & 10 & 15 \\ \hdashline
	$\bsb_{[1]}$& $b_{[1],i}$ &	1 & 3 & 2 & 1 & 3 & 5\\
	$\bsb_{[3]}$& $b_{[3],i}$ &	-3 & -2 & 6 & 1 & 3 & 5\\
	$\bsb_{[6]}$& $b_{[6],i}$ &	-3 & -2 & -1 & -3 & -5 & 15 
\end{tabular}
\end{center}
\caption{Different disaggregation results for $k=1,3,6$ and a specific $\bsa$ of length $n=6$.}
\label{tab_hierarch_example}
\end{table}

In the general $\bsb_{[k]}$ setting, the summation matrix $\bsS_{[k]}$ is given by 
\begin{equation}
	\bsS_{[k]} = \left[\begin{array}{c:c:c} 
		\bsNull_{n-k,k-1} & \bsone_{n-k} &\bsU_{n-k} \\ \hdashline[2pt/2pt] 
\bsL_{k-1} & \bsone_{k-1} &\bsNull_{k-1,n-k} \\ \hdashline[2pt/2pt] 
\multicolumn{3}{c}{\bsI_n}  
\end{array}\right]
\label{eq_summation_matrix_Sk}
\end{equation}
where $\bsL_{k-1}$ is a $(k-1)$-dimensional matrix that contains $1$ on the lower anti-diagonal.
The general summation matrix $\bsS_{[k]}$ in \eqref{eq_summation_matrix_Sk} also nests the canonical case $\bsS$ in \eqref{eq_summation_matrix_S} for $k=1$.
The vector $\bsy_{[k]}$ in the corresponding hierarchy that satisfies $\bsy_{[k]} = \bsS_{[k]} \bsb$ is 
\begin{equation} \bsy_{[k]} = 
	\begin{bmatrix} \bsa_{[-k]} \\ \bsb_{[k]} \end{bmatrix}
	\label{eq_def_yk}
	\end{equation}
 where we define $\bsa_{[-k]}$ as the reversed vector $\bsa$ without the $k$th element, i.e. 
 $\bsa_{[-k]}=(a_n,\ldots, a_{k+1},a_{k-1},\ldots, a_1)'$. 
 Clearly, for $k=1$ we have $\bsy = \bsy_{[k]}$ (see definition 	\eqref{eq_def_yk}).

However, the different representations of the hierarchical structure are actually only formal representations and do not automatically provide different reconciled forecasts by themselves, as shown at the end of the next section.
 To show that this is the case, we observe that the following relations hold:
 \begin{equation}
 a_{i}= \begin{cases}
 \sum_{j=k}^i b_{[k],j}	& \text{, if } i>k \\
 \sum_{j=i}^k b_{[k],j}	& \text{, if } i\le k \\
 \end{cases}
 \label{eq_bk_to_a}
 \end{equation}
 and
  \begin{equation}
 b_{i}= \begin{cases}
 a_1	& \text{, if } i=1 \\
 -b_{[k],i-1}	& \text{, if } 1<i<k+1 \\
 b_{[k],i}	& \text{, if } i\ge k+1. \\
 \end{cases}
 \label{eq_bk_to_b}
 \end{equation}
 
 These relations help us define Equation \eqref{eq_def_bki} in matrix form as $\bsb_{[k]} = \bsA_{[k]}\bsa$ with matrix $\bsA_{[k]}$, which yields $\bsy_{[k]} = \bsS_{[k]}\bsA_{[k]}\bsa$. For $k=1$, we have $\bsA_{[k]} = \bsD_n$ from Equation \eqref{eq_a_to_b}. $\bsA_{[k]}$ has the form:
 
  \begin{equation}
 \bsA_{[k]} = 
 \left[\begin{array}{c:c:c} 
 \bsI_{k-1}+ \begin{bmatrix} \bsnull_{k-2+1} & -\bsI_{k-2} \\ 0 & \bsnull_{k-2+1}' \end{bmatrix} & \begin{bmatrix} \bsnull_{k-2} \\ -1 \end{bmatrix} & \bsNull_{k-1,n-k} \\ \hdashline[2pt/2pt]
  \bsnull_{k-1}' & 1 & \bsnull_{n-k}' \\ \hdashline[2pt/2pt]
  
    \bsNull_{n-k,k-1} & \begin{bmatrix} -1 \\ \bsnull_{n-k-1} \end{bmatrix} & \bsI_{n-k}+ \begin{bmatrix} \bsnull_{n-k-1} & 0 \\ -\bsI_{n-k-1} & \bsnull_{n-k-1}' \end{bmatrix}\\
 \end{array}
 \right].
 \end{equation}
 
In addition, according to definition	\eqref{eq_def_yk},
a matrix $\bsB_{[k]}$ exists that satisfies $\bsy = \bsB_{[k]} \bsy_{[k]}$. 
  We note that $\bsB_{[k]}$ has the structure
 \begin{equation}
 	\bsB_{[k]} = 
	\left[\begin{array}{c:c:c:c:c} 
		\bsI_{n-k} & \bsNull_{n-k,k-1} & \bsNull_{n-k,k-1} &\bsnull_{n-k} & \bsNull_{n-k,n-k} \\ \hdashline[2pt/2pt] 
		\bsnull_{n-k}' & \bsnull_{k-1}' &\bsnull_{k-1}' &1 & \bsnull_{n-k}' \\ \hdashline[2pt/2pt] 
		\bsNull_{k-1,n-k} & \bsI_{k-1} &\bsNull_{k-1,k-1} &\bsnull_{k-1}  & \bsNull_{k-1,n-k} \\ \hdashline[2pt/2pt] 
		\bsNull_{k-1,n-k} & \bsNull_{k-1,k-1} & -\bsI_{k-1} &\bsnull_{k-1}  & \bsNull_{k-1,n-k} \\ \hdashline[2pt/2pt] 
		\bsNull_{n-k,n-k} & \bsNull_{n-k, k-1} &  \bsNull_{n-k, k-1} &\bsnull_{n-k} & \bsI_{n-k} \\ \hdashline[2pt/2pt] 
	\end{array}
 \right].
 \end{equation}
Furthermore, it is easy to check that  $\bsB_{[k]}$ is orthogonal, i.e. it holds $\bsB_{[k]}^{-1} = \bsB_{[k]}'$. %
We observe that $\bsB_{[k]}$ is a generalized permutation matrix that contains permutations and reflection components.

Finally, using $\bsy = \bsB_{[k]} \bsy_{[k]}$, $\bsy_{[k]} = \bsS_{[k]}\bsA_{[k]}\bsa$ and $\bsa = \bsD_n^{-1}\bsb$ we obtain that
 $\bsy = \bsB_{[k]}\bsS_{[k]}\bsA_{[k]} \bsD_n^{-1}\bsb $. Thus, it holds that $\bsS  = \bsB_{[k]}\bsS_{[k]}\bsA_{[k]} \bsD_n^{-1}$ for all $k$, which is valuable for further analysis.
}

\section{Reconciliation approaches}
\label{sec:reconciliation}

In general, the hierarchical structure does not hold if each time series is forecasted individually. As stated by \citet{wickramasuriya2019optimal}, we call these individual forecasts \emph{incoherent} or \emph{base forecasts}, and denote them by $\what \bsy$. \FZ{ The \emph{coherent} or \emph{reconciled forecasts} for which the structure in Figure \ref{fig:hierarch} holds are denoted by $\wtilde{\bsy}$, and they formally satisfy: $\wtilde{\bsy} = \bsS \wtilde{\bsb}$, where $\wtilde{\bsb}$ are the bottom values in $\wtilde{\bsy}$.} %

The coherency of forecasts is a desired property of aggregated curves for further analysis, e.g. for trading applications, but the main application of reconciliation is generating accurate values for the aggregated level $\bsa$. 
All linear reconciliation approaches for any hierarchical structure can be compactly written in matrix notation as
\begin{equation}
\wtilde{\bsy} = \bsS \bsP \what{\bsy}
\label{eq:hierarchmatrix}
\end{equation}
\FZ{The $ (2n-1) \times n$-dimensional \emph{mapping matrix} $\bsP$ maps the base forecasts $\bsy$ to the bottom level, \citep{hyndman2018forecasting}. %
 $\bsP$ is different for each reconciliation approach. The product $\bsS\bsP$ is sometimes referred to as the \emph{reconciliation} or \emph{projection matrix}.

In the previous section, we described the canonical representation and other representations. We note that they imply the same hierarchical structure and it holds that $\bsS  = \bsB_{[k]}\bsS_{[k]}\bsA_{[k]} \bsD_n^{-1}$. Furthermore, the results are actually equivalent for specific reconciliation approaches considered in the simulation study and application, i.e. they are not affected by the choice of $k$ (see Section \eqref{subsec_other_repre_rec} for more details). Therefore, in the following subsections, we only consider the canonical representation.}

\subsection{Simple benchmark approaches}
\PG{Table \ref{tab:recon_bench} summarizes the formulas for our three benchmark reconciliation approaches: bottom-up, top-down and aggregated-down. The first two approaches are well-established procedures, so we will not provide their details but instead we refer to previous studies (\cite{gross1990disaggregation,hyndman2018forecasting}). For the top-down  and aggregated down methods, we provide three options for calculating the disaggregating proportions, i.e., using the average ratio (ar), ratio of averages (ra), and forecasted values (fo). Many more options could be considered for the top-down approach (\cite{gross1990disaggregation}). 
We describe the new aggregated-down approach in the next subsection.}

\begin{table}[ht]
	\centering
	\scalebox{1}{
		\begin{tabular}{|l|l|}
			\hline
			\textbf{Method} & \multicolumn{1}{c|}{\textbf{Mapping matrix P}} \\ \hline
			Bottom-up & $\bsP_{\textrm{bu}} = \begin{bmatrix} \bsNull_{n\times (n-1)} & \bsI_n \end{bmatrix}$  \\ \hline
			Top-down & 
			\begin{tabular}{l} 
				$\bsP_{\text{td}} = \begin{bmatrix} \bsp & \bsNull_{ n\times (2n -2)} \end{bmatrix}$  \\
				$\bsp = ( p_{1} , \dots ,  p_{n} )'$ \\ 
				Average ratio:  $\what p_{\text{ar},j} = \frac{1}{T} \sum_{t=1}^{T}{\frac{b_{j,t}}{a_{n,t}}}$ \\ 
				Ratio of averages: 	$\what p_{\text{ra},j} = \frac{\frac{1}{T} \sum_{t=1}^{T}{b_{j,t}}}{\frac{1}{T} \sum_{t=1}^{T}{a_{n,t}}}$ \\ 
				Forecasted values: \\
				$
				\what p_{\text{fo},j} =  
				\begin{cases}
				\frac{\what b_{n}}{\what a_{n-1} + \what b_n}, & \text{for } j = n \\
				\frac{\what b_j}{\what a_{j-1} + \what b_j} 
				\prod_{i=j}^{n-1} \left (\frac{ \what a_i}{\what a_i + b_{i+1}} \right ) & \text{for } 1 < j < n, \\
				\prod_{i=1}^{n-1} \left (\frac{ \what a_i}{\what a_i + b_{i+1}} \right ) & \text{for } j = 1.
				\end{cases}
				$
			\end{tabular} \\ \hline
			Aggregated-down & 
			\begin{tabular}{l} 
				$ \bsP_{\textrm{ad}} = \begin{bmatrix} \bsQ_{n \times n} & \bsNull_{n \times (n-1) } \end{bmatrix} $ \\
				$\bsQ_{n \times n} = \textrm{Antidiag}(\textbf q) = \textrm{Antidiag}((q_1, \dots q_n)')$ \\
				$q_1=1$ \\ 
				Average ratio: $\what{q}_{\text{ar}, j} = \frac{1}{T} \sum_{t=1}^{T}{\frac{a_{j,t} - a_{j-1,t}}{a_{j,t}}} = \frac{1}{T} \sum_{t=1}^{T}{\frac{b_{j,t}}{a_{j,t}}}$ \\ 
				Ratio of averages: $\what q_{\text{ra},j} = \frac{\frac{1}{T} \sum_{t=1}^{T} a_{j,t} - a_{j-1,t} }{\frac{1}{T} \sum_{t=1}^{T}a_{j,t}} = \frac{\frac{1}{T}\sum_{t=1}^{T} b_{j,t} }{ \frac{1}{T}\sum_{t=1}^{T} a_{j,t}}	$ \\ 
				Forecasted values: $\what{q}_{\text{fo},j} = \frac{\what a_{j} - \what a_{j-1}}{ \what a_{j}} \textrm{ for } j > 1$, 
			\end{tabular} \\ \hline	
		\end{tabular}
	}
	\caption{Reconciliation methods and their formulas for aggregated curves ($\bsNull_{\cdot \times \cdot}$ is a zero matrix of indicated dimensions)}
	\label{tab:recon_bench}
\end{table}

We use \textit{bu} as an abbreviation for bottom-up, and \textit{tdar}, \textit{tdra}, \textit{tdfo} for the top-down approaches using the average ratio, the ratio of averages, and forecasted values respectively.

\subsection{Aggregated-down approach}
\label{sec:ad}
\PG{The \emph{aggregated-down} approach is essentially a localized top-down approach, where the disaggregating proportions are calculated based on the node above and not based on the top-most aggregated level.
The motivation for this approach is based on the simple assumption that proportions calculated using values that are closer in the hierarchical structure are expected to be more accurate than when using values that are further away. We provide support for this assumption based on our simulation study in the next section.}

 \PG{We denote the corresponding disaggregating proportions by $q_j$ 
where $\bsq = ( q_1 , \dots , q_{n} )'$ is the vector of proportions.
By design (see Fig. \ref{fig:hierarch}), it holds that
\begin{equation}
	b_j = q_j a_{n-j+1}.
\end{equation}
Thus, $q_j$ is the disaggregation proportion for $b_j$ given $a_{n-j+1}$ at the corresponding connection of the tree.
	}
The mapping matrix is defined as
$$\bsP_{\textrm{ad}} = \begin{bmatrix} \bsQ & %
\bsNull_{n \times (n-1) } \end{bmatrix},$$
where $\bsQ = \text{Antidiag}( \bsq)$ is a $n \times n$ anti-diagonal matrix with the elements on the anti-diagonal starting from top to bottom, i.e., 
 $$\bsQ = 
 \begin{bmatrix} 
  0 & \dots & 0  & q_1   \\ 
 0 &   \dots & q_2 & 0   \\ 
 \vdots & \vdots   & \ddots & \vdots \\ 
 q_n & 0   & \dots  &  0 \\ 
 \end{bmatrix} .$$

We require that $q_1=1$ every time, which corresponds to $\tilde{b}_1 = \hat{b}_1$.
\PG{Analogous to the top-down approach, the proportions can be estimated using various methods, which are summarized in Table \ref{tab:recon_bench}. It should be noted that the methods for calculating the average ratio and ratio of averages proportion calculation methods have similar complexity for the top-down and aggregated-down approaches. However, the formula for calculating the aggregated-down approach using the forecasted values $\what{\bsy}$ is much simpler than that for the top-down approach.
}

We use \textit{adar}, \textit{adra}, \textit{adfo} as abbreviations for the aggregated-down approaches using the average ratio, the ratio of averages, and forecasted values, respectively (see Table \ref{tab:recon_bench}).

\subsection{Optimal reconciliation approach}
\label{sec:opt}
\citet{hyndman2018forecasting} and \citet{wickramasuriya2019optimal} introduced the optimal reconciliation approach by showing that the optimal mapping matrix that obtains the best, unbiased coherent forecasts is given by
\begin{equation}
	\bsP_{\textrm{op}} = (\bsS'\bsW^{-1}\bsS)^{-1}\bsS' \bsW^{-1}
	\label{eq:G}
\end{equation}
where $\bsW=\text{Var}[\bsy-\what \bsy]$ is the variance-covariance matrix of the base forecast errors. Equation (\ref{eq:G}) is the result obtained by minimizing the variance of the coherent forecasts. $\bsW$ is not known \FZ{and must be estimated. We consider multiple estimators for $\bsW$,  which %
 are listed below.} The first five estimators were proposed by \citet{hyndman2018forecasting} and \citet{wickramasuriya2019optimal}.
\begin{enumerate}
	
	\item $\bsW_{\text{opols}}=\bsI_{2n-1}$, where $\bsI_{2n-1}$ is the identity matrix. \FZ{For the optimal projection matrix $\bsP$ in the aggregated curves setting, we have
	$ \bsP_{\text{opols}} = (\bsS'\bsS)^{-1}\bsS'$ .
	We note that $\bsS'\bsS$ is of full rank and invertible because it holds 
$$\bsS'\bsS 
= [\bsone_{n-1}:\bsU_{n-1}]'[\bsone_{n-1}:\bsU_{n-1}] + \bsI_n = 
	 \begin{bmatrix} 
	n & n-1 & n-2 &\cdots & 2 & 1 \\
	n-1 & n & n-2 &\cdots & 2 & 1 \\
	n-2 & n-2 & n-1 &\cdots & 2 & 1 \\
	\vdots & \vdots& \vdots &\ddots & \vdots & \vdots \\
	2 & 2 &2 & \cdots & 3 & 1 \\
	1 & 1 & 1&\cdots & 1 & 2 \\
	 \end{bmatrix} .
$$
Thus, we obtain
$ \bsP_{\text{opols}} = 
	\left( [\bsone_{n-1}:\bsU_{n-1}]'[\bsone_{n-1}:\bsU_{n-1}] + \bsI_n\right)^{-1} \bsS'$.
	However, the assumption that
	$\text{Var}[\bsy-\what \bsy]$ is constant is usually not realistic in an aggregated curve setting because we expect a tendency towards larger variances for higher aggregation levels.} We use \textit{opols} as an abbreviation for this approach.

	\item $\bsW_{\text{oplambda}}= \bf \Lambda$,  $\bf \Lambda = \text{Diag}(\bsS \bsone_n)$, where $\bsS$ is the summation matrix and $\bsone_n$ is a unit vector of the same dimension as the number of bottom-level time series. In our aggregated curves setting, it holds that $\text{Diag}({\bf \Lambda})=\bsS \bsone_n=(n,n-1,\dots,2,1,1,\dots,1)'$, \FZ{which means that the higher aggregated values of the curves are weighted less, proportional to their level of aggregation starting from $1$ as the lowest and $n$ as the top. This corresponds to a setting where all bottom-level forecasts $\bsb$ have the same variance and are uncorrelated.} We use \textit{oplambda} as an abbreviation for this approach.

	\item $\bsW_{\text{opwls}}= \what \bsW_{\text{dcov}}$, where $ \what \bsW_{\text{dcov}}$ is an estimator for $ \bsW_{\text{dcov}}= \text{Diag}( \bsW_{\text{cov}})$ with $\bsW_{\text{cov}}$ as the covariance matrix of the errors associated with $\bsy$.
	$\bsW_{\text{cov}}$ can be estimated by the sample covariance $\what \bsW_{\text{cov}}$
	, i.e.
	$\what{\bsW}_{\text{cov}} = \frac{1}{n} \textbf E' \textbf E = \frac{1}{n} \sum_{i=1}^n \bse_i^{\space} \bse_i'$ and $\bsE$ is the matrix of residuals generated by and arranged in the same order as the base forecasts. 
	\FZ{ The $\bsW_{\text{opwls}}$ approach may be regarded as a generalization of the $\bsW_{\text{oplambda}}$ approach. This design corresponds to a setting where the individual forecast errors have different variances but are uncorrelated.} We use \textit{opwls} as an abbreviation for this approach.
	
	\item $\bsW_{\text{opcov}}= \what \bsW_{\text{cov}}$, where $\bsW_{\text{cov}}$ is the full sample covariance matrix of the error terms. The underlying setting corresponds to a situation where the forecast errors have varying variances and exhibit linear dependence. We use \textit{opcov} as an abbreviation for this approach.
	
	\item $\bsW_{\text{opshrink}} = \lambda\what \bsW_{\text{dcov}} + (1 - \lambda)\what \bsW_{\text{cov}}$, 
	where $\lambda$ is the shrinkage intensity parameter. \citet{schafer2005shrinkage} proposed setting $$\lambda = \frac{\sum_{i \ne j} \what{\text{Var}}(\what r_{ij})}{\sum_{i \ne j}\what r_{ij}^2},$$
	where $\what r_{ij}$ is the $ij$th element of the 1-step-ahead sample correlation matrix. The authors implemented the formulas in the \texttt{corpcor} R package. We use \textit{opshrink} as an abbreviation for this approach.
	
	\item Ledoit-Wolf covariance matrix estimator with shrinkage toward constant correlation \citep{ledoit2004honey}: 
	$\bsW_{\text{opledoitwolf}} =  \delta \textbf F + (1-  \delta )\what \bsW_{\text{cov}}$, where $\what \bsW_{\text{cov}}$ is the sample covariance matrix,  $\textbf F$ is the shrinkage target with constant correlation defined with element $f_{ij} = \bar r \sqrt{\what w_{ii} \what w_{jj}}$ on the $i$th row and $j$th column, $\what w_{ij}$ is the corresponding element of $\what \bsW_{\text{cov}}$, and
	$$\bar r = \frac{2}{(N-1)N} \sum_{i=1}^{N-1} \sum_{j=i+1}^N r_{ij},$$ $r_{ij}=\frac{\what w_{ij}}{\sqrt{\what w_{ii},\what w_{jj}}}$. The estimator of the shrinkage intensity $\delta$ was introduced by \citep{ledoit2004honey}. We use \textit{opledoitwolf} as an abbreviation for this approach. %
		
	\item Covariance matrix estimation using graphical lasso \citep{friedman2008graphlasso}: 
	$$\bsW_{\text{opglasso}} = 
	\begin{bmatrix} 
	\bsW_{11} & \bsw_{12} \\
	\bsw'_{12}       & w_{22}       \\
	\end{bmatrix},$$ 
	where $\bsW_{\text{opglasso}}$ is partitioned as shown and estimated by using coordinated descent to solve
	$$\bsbeta_{\text{opglasso}}= \argmin_{\bsbeta} \left\{ \frac{1}{2} \| \bsW_{11}^{1/2} \bsbeta - \bsb \|^2 + \rho \| \bsbeta \|_1 \right\},$$
	where $\bsb = \bsW_{11}^{-1/2} \what \bsw_{12}$. The optimal $\bsbeta$ is used to generate the optimal $\bsw_{12} = \bsW_{11} \bsbeta_{\text{opglasso}}$. Initially, $\bsW_{\text{opglasso}}$ is set to $\bsW =  \what{\bsW}_{\text{cov}}+ \rho \bsI_n$. We used the algorithm as implemented by \citep{friedman2008graphlasso} in the \texttt{glasso} R package. We use \textit{opglasso} as an abbreviation for this approach.
	
\end{enumerate}

\FZ{
\subsection{Optimal reconciliation for other representations}
\label{subsec_other_repre_rec}
Next, we briefly describe the optimal reconciliation approach for other representations. In this situation, for some $k$, we have
$$\wtilde{\bsP}_{[k]} = (\bsS_{[k]}'\bsW_{[k]}^{-1}\bsS_{[k]})^{-1}\bsS_{[k]}' \bsW_{[k]}^{-1}.$$
Furthermore, we can show that under mild assumptions regarding the forecast method and the reconciling matrix $\bsW_{[k]}$, the reconciliation approach preserves the hierarchical structure, which holds in the sense that the result does not depend on the choice of $k$: 
\begin{theorem}
	If $\what{\bsy} = \bsB_{[k]} \what{\bsy}_{[k]}$ and $\bsW^{-1}_{[k]} = \bsB_{[k]}'\bsW^{-1}\bsB_{[k]} $ then it holds that
	$$\wtilde{\bsy} =  \bsB_{[k]} \wtilde{\bsy}_{[k]}.$$
\end{theorem}	
The proof is presented in the \hyperref[sec:appendix]{Appendix}. We recall that $\bsB_{[k]}$ is orthogonal, and thus the assumption
$\what{\bsy} = \bsB_{[k]} \what{\bsy}_{[k]}$ is satisfied if the forecasting algorithm that provides $\what{\bsy}$ is invariant to orthogonal transformations, e.g., this holds for linear regressions.
In addition, we note that 
$\bsW^{-1}_{[k]} = \bsB_{[k]}'\bsW^{-1}\bsB_{[k]} $ is trivially satisfied if $\bsW = \bsI_{2n-1}$ due to orthogonality of $\bsB_{[k]}$.

}

\section{Simulation study }
\label{sec:simulation}

\PG{The new aggregated-down approach is motivated based on the simple idea of closeness, i.e. proportions calculated using values that are closer in the hierarchical structure are expected to be more accurate than those calculated when using values that are further away due to avoidance of error aggregation. To support this claim, we conducted a simulation study, which also allowed us to compare the performance with other approaches.}

\subsection{Study design}

\PG{We simulated Vector Auto-Regressive VAR(1) processes for the bottom-level values. We replicated the simulation 1000 times for each combination of the parameters specified in Table \ref{tab:simSetup}.}

\begin{table}[ht]
	\centering
	\scalebox{0.9}{
		\begin{tabular}{|l|l|}
			\hline
			Number of historical observations ($N$)	    & $N\in \{16, 64, 256\}$ \\
			Number of bottom-level values ($n$) 		& $n\in\{4, 16, 64\}$ \\
			Coefficient matrix ($\bsPhi$) 				& $\bsPhi =  a \bsI_n$ for $a\in\{0.2,0.5,0.7,0.95\}$ \\
			Error variance-covariance matrix ($A$)		& $A= \bsI_n $ \\
			\hline
		\end{tabular}
	}
	\caption{Simulation study setup: combinations for a VAR(1) process. }
	\label{tab:simSetup}
\end{table}

\PG{We also considered the case of correlated errors with the variance-covariance matrix $A= 0.3 \bsI_n + 0.7 \bsone \bsone'$ in combination with the coefficient matrix $\bsPhi =  0.7 \bsI_n$.  We generated the aggregated values from the bottom-level simulations and fitted an Auto-Regressive AR(1) process without intercept for each level of the hierarchy. We calculated the forecast accuracy from 1-step-ahead forecasts for each level. Next, we computed the root mean square errors (RMSE) value of each reconciling approach considered.}

\subsection{Study results}

\PGG{
The results are shown in Table \ref{tab:sim07} for the setup with $\bsPhi = 0.7 \bsI_n$ and $A= \bsI_n$. We left out the results for the average ratio and ratio of averages approaches for top-down and aggregated-down since they were always greatly inferior to those obtained using the base case. We also omit some optimal methods for brevity because these tended to produce very similar results. }

\begin{table}[ht]
	\centering
	\scalebox{0.7}{
		\begin{tabular}{|r|rrr|rrr|rrr|}
			\hline
			\textbf{No. of bottom-level values ($n$)} & \multicolumn{3}{c|}{\textbf{4}} & \multicolumn{3}{c|}{\textbf{16}} & \multicolumn{3}{c|}{\textbf{64}}\\ \hline
			\textbf{No. of observations ($N$)} & \textbf{16} & \textbf{64} & \textbf{256} & \textbf{16} & \textbf{64} & \textbf{256} & \textbf{16} & \textbf{64} & \textbf{256} \\ 
			\hline
			base & 1.40 & 1.39 & 1.32 & 2.24 & 2.27 & 2.24 & 4.33 & 4.13 & 4.15 \\ 
			bu & 1.40 & 1.39 & 1.32 & 2.25 & 2.26 & 2.24 & 4.32 & 4.12 & 4.15 \\ 
			tdfo & 18.0 & 8.07 & 1.61 & 55.7 & 99.5 & 74.1 & 24200 & 1610 & 605 \\ 
			adfo & 1.42 & 1.39 & 1.32 & 2.26 & 2.27 & 2.24 & 4.35 & 4.13 & 4.15 \\ 
			opols & 1.39 & 1.39 & 1.32 & 2.23 & 2.27 & 2.24 & 4.33 & 4.12 & 4.15 \\ 
			opwls & 1.39 & 1.39 & 1.32 & 2.22 & 2.26 & 2.24 & 4.30 & 4.12 & 4.15 \\ 
			oplambda & 1.39 & 1.39 & 1.32 & 2.21 & 2.26 & 2.24 & 4.29 & 4.12 & 4.15 \\ 
			opshrink & 1.39 & 1.39 & 1.32 & 2.23 & 2.27 & 2.24 & 4.35 & 4.13 & 4.15 \\ 
			\hline
		\end{tabular}
	}
	\caption{Simulation study RMSE results for the VAR(1) setup $\Phi = 0.7\bsI_n$ and $A=\bsI_n$ for selected reconciliation methods}
	\label{tab:sim07}
\end{table}

\PGG{
	The key finding from this simulation study is that the aggregated-down approach using forecasted values was very similar to the other methods, and even optimal ones. By contrast, the top-down approach using forecasted values was markedly inferior in the setups considered. As expected, the accuracy generally decreases as more bottom-levels ($n$) were added. However, this was disproportionally the case with the top-down approach, which was caused by  extreme sensitivity to outliers because the proportions were multiplicatively connected and common factors were present across the hierarchy. We refer to this issue as the \textit{error inheritance} of proportions.
	
	 This situation is clear when we consider the formulas. For illustrative purposes, we consider the simplest proportions $\hat p_{fo,n}$, and $\hat p_{fo,n-1}$:
	 $$\hat p_{fo,n} = \frac{\hat b_{n}}{\hat a_{n-1}+ \hat b_n}, \ \ \hat p_{fo,n-1} = \frac{\hat b_{n-1}}{\hat a_{n-2}+ \hat b_{n-1}} \underbrace{\frac{\hat a_{n-1}}{ \hat a_{n-1}+ \hat b_n}}_{=1-\hat p_{fo,n}} $$
	 Clearly, the formulas are sensitive to outliers when the denominator $\hat a_{n-1}+ \hat b_n$ is very small, i.e. when $\hat a_{n-1} \approx -\hat b_n$. If this is the case, then all of the subsequent proportions will generally be outliers because $\hat p_{fo,n}$ is part of $\hat p_{fo,n-1}$, and so on. It should be noted that this occurs if we allow negative values. If we only consider monotonously increasing curves, then these outliers do not occur, as shown in Table \ref{tab:sim07abs}, where we simulated a strictly positive AR(1) process leading to errors close to the other methods.
	 To understand the sensitivity to outliers, Table \ref{tab:sim07nooutliers} shows the results for the same setup with all outliers defined as $|\hat p_{fo,1}|>50$ removed. The errors improved for greatly \textit{tdfo}, especially with increasing values of $N$.	 
}

\PG{
This issue is essentially error aggregation and is important considering the specific hierarchical structure of aggregated curves because the hierarchy considered is completely vertical and each additional point on the curve deepens the hierarchy instead of widening it. Thus, if we allow for negative values, the outliers become larger for \textit{tdfo} when the hierarchy is deeper. The aggregated-down approach is not sensitive to this distance issue since proportions are not multiplicative, but instead they are always calculated using values immediately above the hierarchy level considered, and non-monotonic curves can be easily accommodated.} 
The other simulation setups yielded similar results, as shown in Appendix Tables \ref{tab:sim05}, \ref{tab:sim02}, \ref{tab:sim95}, and \ref{tab:simcorr}.

It is important to note that we consider the aggregated-down approach as a simple benchmark approach. It is not equivalent to more sophisticated optimal reconciliation methods. Nevertheless, aggregated-down has the advantages of being intuitive and using very simple formulas for calculating proportion, especially compared to the top-down approaches, and is programmatically simple as well as efficient. Furthermore, it is not affected by the error aggregating issue, unlike the top-down approach.

\section{Application}
\label{sec:Application}
\subsection{Supply and demand curves on day-ahead electricity markets}

\begin{figure}[h!]
	\centering
	\includegraphics[width=1\textwidth]{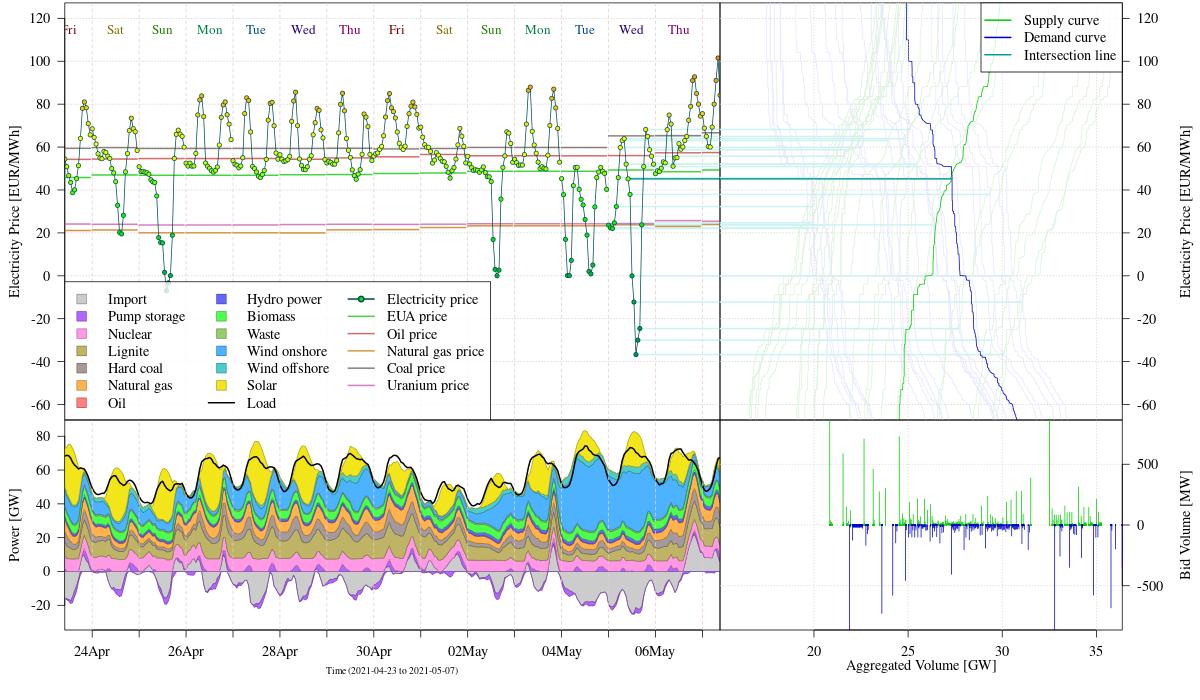}
	\caption{German-Luxembourg day-ahead market: Hourly day-ahead electricity price time series with relevant commodity price time series (top left) with
	corresponding 24 supply and demand curves on the price grid between -60 and 120 EUR/MWh for 5 May 2020 with highlighted curves for
	11:00 (top right), power generation, import and consumption time series (bottom left), and bid structure of 5 May
	2020 11:00 (bottom right).}
	\label{fig:damarket}
\end{figure}

\FZ{   The European day-ahead electricity market is a daily, blind auction market where electricity prices are determined for each hour of the next day. The auction closes at 12:00 CET each day until when the market participants, i.e. the buyers and sellers, can submit buy (bid) and sell (ask) orders for each hour of the next day. These bids are the volumes that buyers are willing to buy and that producers are willing to sell at certain prices. For the time range of data considered, bid volumes can be specified for prices ranging from -500 EUR/MWh up to 3000 EUR/MWh with an increment of 0.1 EUR/MWh\footnote{Since 2022-05-11 the upper price limit was increased to 4000 EUR/MWh.}. 
After gate closure, the volumes bid across all participating countries are aggregated and unique prices and volumes are generated for each separate market area. One of these areas is Germany-Luxembourg, which we modeled and forecasted in this study. The market clearing results as well as the 24 supply and demand curves are published shortly after gate closure at 12:00 CET. All major relevant aspects of the German day-ahead electricity market are illustrated in Figure \ref{fig:damarket} (see \cite{petropoulos2022forecasting} for more details). Only selected prices can be bid on out of a total of 35001 possible prices, so the curves have a characteristic step-like appearance due to the many zero-volume prices, i.e. prices that were not bid on \cite{xmodel}. }

\begin{figure}[h!]
	\centering
	\subfigure[Volumes bid on all prices]{\includegraphics[width=\textwidth, height=0.3 \textheight]{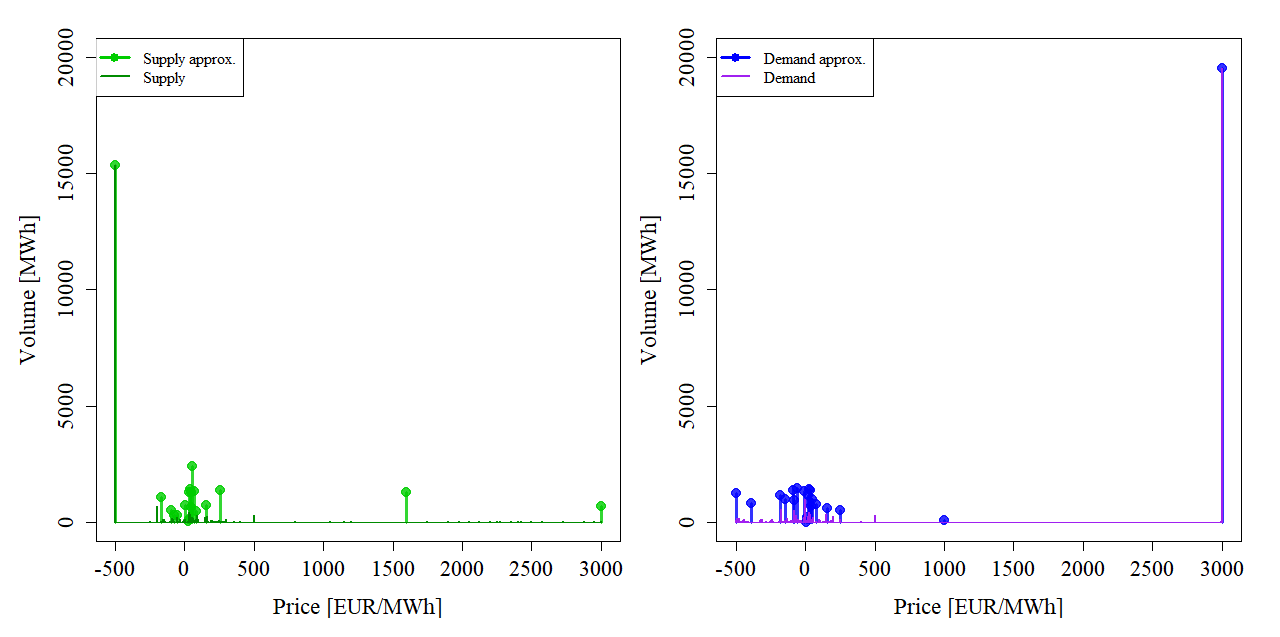}}
	\subfigure[Volumes bid on selected price range]{\includegraphics[width=\textwidth, height=0.3 \textheight]{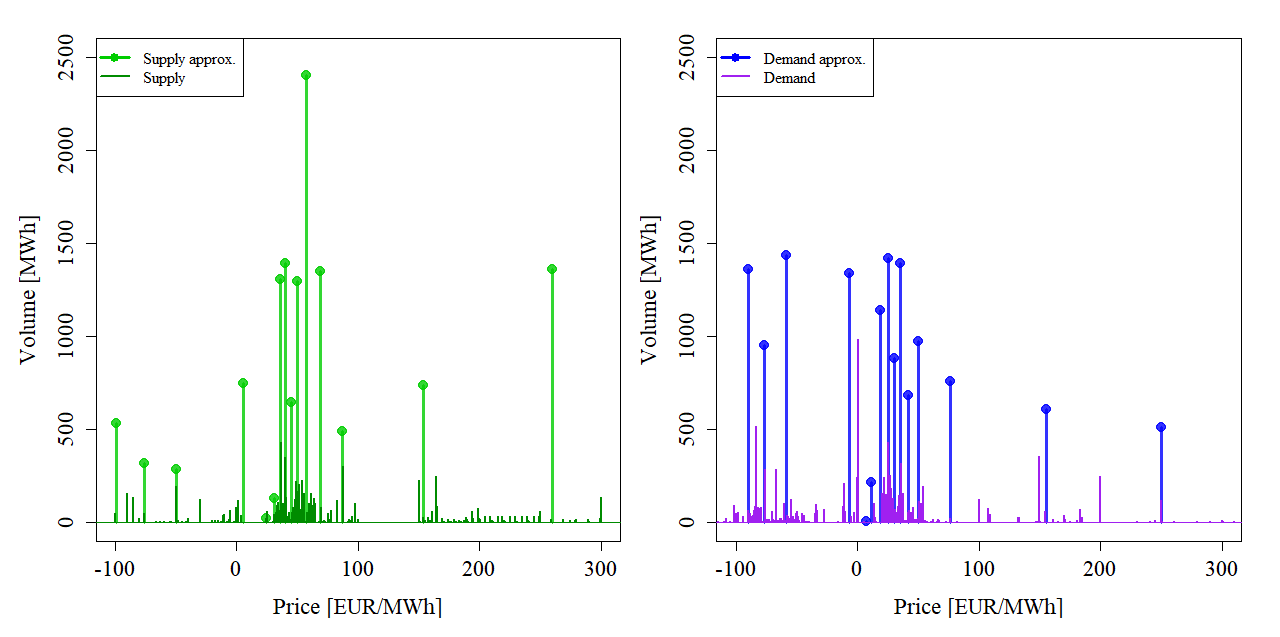}}
	\caption{Volume bid per price for 2019-03-01 12:00:00, i.e. marginal values of the Supply and Demand curves. Volumes bid per price class are also shown, i.e. the marginal values of the Supply and Demand approximations. The bottom plots represent a magnification around $0$.}
	\label{fig:curve_marg}
\end{figure}

\begin{figure}[h!]
	\centering
	\includegraphics[width = \textwidth,height= \textheight, keepaspectratio]{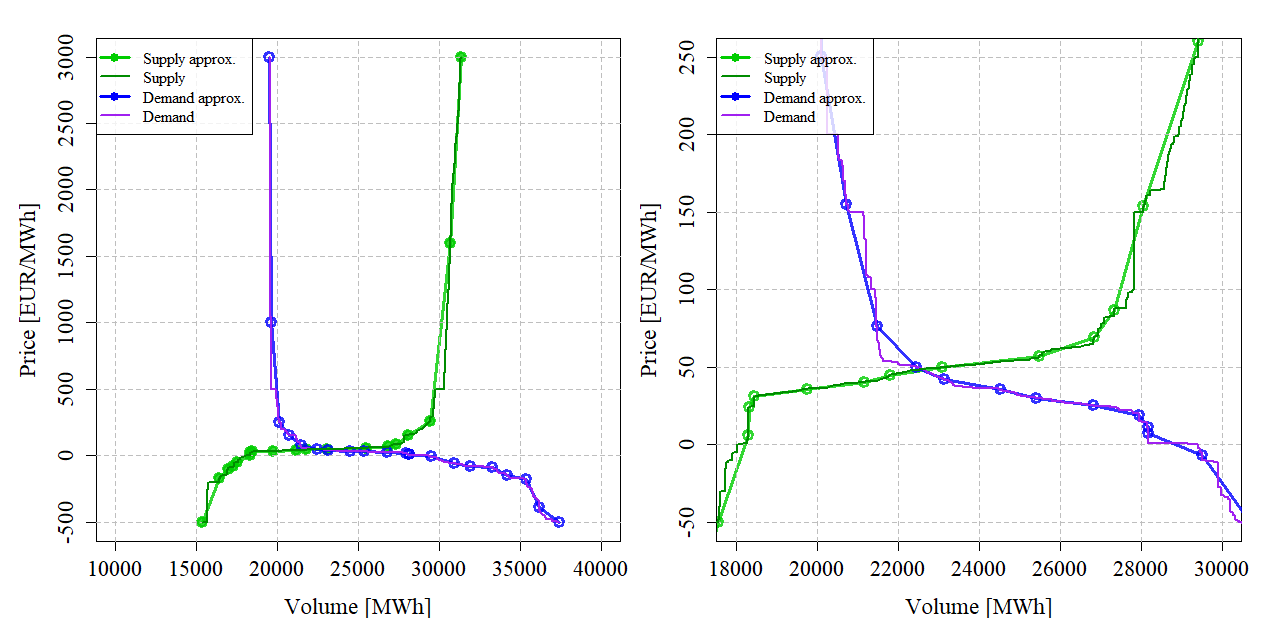}
	\caption{Supply and Demand curves for 2019-03-01 12:00:00, created by the cumulative aggregation of marginal volumes from Figure \ref{fig:curve_marg}. Approximations of the Supply and Demand curves, created by cumulative aggregation of the price classes, are also shown. The plot on the right represents a magnification around the intersection point.}
	\label{fig:curveapprox}
\end{figure}

Figure \ref{fig:curve_marg} shows the German volumes bid for each price for a selected day and hour. In total, there are 35001 prices that can be bid on for each hour, where most usually have a volume of zero, \PG{as can be observed by the many empty intervals between the bars}. Certain prices are generally preferred by market participants, such as multiples of 10 and those in the vicinity of 40 EUR/MWh, as shown by the clusters formed around these prices in Figure \ref{fig:curve_marg}. The hourly bids are aggregated to produce the day-ahead supply and demand curves for each hour, as shown in Figure \ref{fig:curveapprox}.

\subsection{The X-Model}
\label{sec:xmodel}

The X-Model introduced by \citep{xmodel} is an approach for forecasting day-ahead electricity prices as the intersection of supply and demand curves. The electricity prices are not modeled directly but instead they are obtained from the forecasts of the whole day-ahead curves as their intersection.

\citep{xmodel,ziel2018probabilistic,kulakov2020x, haben2021probabilistic} modeled the day-ahead supply and demand curves by grouping the prices into price classes, thus drastically reducing the dimensionality, e.g., from 350001 prices to a few price classes (40 in our study), to make the problem computationally feasible. The prices are split into price classes by inverting the supply and demand curves at a pre-specified grid of equidistant volumes \citep{xmodel}. The result is shown in Figure \ref{fig:curveapprox}, where the curves are approximated by ca. 20 points each. \PG{This form of dimensionality reduction via binning can be interpreted as a special case of applying FDA where the basis functions are constants, i.e. dummy variables, that do not overlap. This works very well in our case because the original curves are not smooth. An additional advantage is computability because it would be quite costly to smoothen the curves first, model them, and then un-smooth them again for the final representation. \citep{xmodel} use a simple but very effective method for reconstructing the forecasted curves from the price class approximations, i.e., by using historical proportions for each price.  This method was shown to capture the shape of the original curves very well without requiring vast computational resources. However, this reconstruction step is only important for calculating the price as the intersection of the supply and demand curves and does not affect the forecasting accuracy of the curves themselves, hence we refer to the original X-model paper for more information \citep{xmodel} as well as other papers that use more classical FDA approaches to implement the X-model \citep{soloviova2019modeling, soloviova2021efficient}.}

Thus, the response variable is the sum of the volume within a price class, which is modeled separately for every hour and class. For $40$ price classes comprising both supply and demand, a total of $40 \times 24$ regressions must be conducted to forecast curves for all hours in the next day. \citep{xmodel} modeled each price class volume marginally, and thus the supply and demand curves were generated by cumulatively summing up the forecasted values, which is equivalent to forecasting only the bottom-level values and using the bottom-up reconciliation approach described earlier. In this study, we modeled both the marginal and cumulative responses, and used them to compare and contrast the different reconciliation approaches. It should be noted that the method using marginal values will always be equivalent to the bottom-up approach.

To model the responses, we used a combination of autoregressive and external regressors.  Let then $X_{S,d,h}^{(c)}$ and $X_{D,d,h}^{(c)}$ be the supply and demand volumes at day $d$ and hour $h$ of price class $c$ among the price classes generated for the supply and demand curves, respectively. These constitute the bulk of the regressor matrix because each price class volume will depend on its lags according to a specific lag structure. The external regressors are denoted by $X_{X,d,h}^{(1)}, \dots, X_{X,d,h}^{(M_X)}$ for a total of $M_X$ external regressors, and they comprise the prices for coal, gas, oil and CO$_2$ emissions (EUAs), the day-ahead prices and volumes on the previous day, as well as the day-ahead forecasts for the country-wide load, solar, onshore wind, and offshore wind production. \PG{The day-ahead data were taken from \textit{www.epexspot.com} and the forecasts from \textit{www.entsoe.eu}.}

Let $M_S$ and $M_D$ be the number of price classes for the supply and demand curves, respectively. Then. all regressors can be compactly written as 
$$\textbf{X}_{d,h} = \left ( X_{1,d,h}, \dots, X_{M,d,h} \right )' = \left ( \left (X_{S,d,h}^{(c \in C_S)} \right ), \left ( X_{D,d,h}^{(c \in C_D)} \right ), \left ( X_{X,d,h}^{(c \in C_X)} \right ) \right )',$$ 
where $C_S$ and $C_D$ are the sets of price classes for the supply and demand sides, respectively, $C_X$ is the  set of external regressors $C_X = \left \{ X_{X,d,h}^{(1)} 1, \dots, X_{X,d,h}^{(M_X)} \right \}$, and $M=M_S+M_D+M_X$.

To capture the weekly seasonality, we also included dummy regressors for every day of the week which we denoted by a function $W_k(d)$ that returns the day of the week $d$. The full model can be written as

$$X_{m,d,h} = \sum_{k=1}^{M} \sum_{j=1}^{24} \sum_{k \in \mathcal{I}_{m,h}(l,j)} \phi_{m,h,l,j,k} X_{l,d-k,j} + \sum_{k=2}^{7} \psi_{m,h,k} W_k(d) + \varepsilon_{m,d,h},$$

for $m \in \{1, \dots, M_S + M_D\}$ and $\mathcal{I}_{m,h}(l,j)$ represents the sets of possible lags, which we defined as
\[
\mathcal{I}_{m,h}(l,j) =  
\begin{cases}
\{ 1, \dots, 30\}, \text{for } m = l \text{ and } h = j\\
\{ 1, \dots, 8\}, \text{ for } (m = l \text{ and } h \ne j) \text{ or } (m \ne l \text{ and } h = j)\\
\{ 1 \}, \text{ for } m \ne l \text{ and } h \ne j
\end{cases}.
\]
We fitted the models using lasso \citep{xmodel} as implemented in the \textbf{glmnet} R package.

\subsection{Data and results}
\label{sec:results}
We conducted a day-ahead rolling window forecasting study for each day between 2019-01-01 and 2019-06-30. We used a rolling window length of $730 = 2 \times 365$ days to forecast each day. The price classes were generated only once for the first day-ahead forecast, i.e. using the 2017 and 2018 data, and they were kept constant throughout the study. All data were hourly except for coal, gas, oil, and EUA prices, which were daily. We used marginal values as regressors to forecast the marginal values $\what b_i$ and cumulative values as regressors to forecast the cumulative values $\what a_i$. We chose $\what a_n \leftarrow \what b_1$ for each day and hour, as in Figure \ref{fig:hierarch}. 

To measure the forecasting accuracy, we used two popular error measures comprising the mean absolute errors (MAE) defined as
$$\textrm{MAE}_m^{\textrm{test}} = \frac{1}{24\cdot \#(\mathcal{D})} \sum_{d\in \mathcal{D}} \sum_{h=0}^{23} |X_{m,d,h}-\what X_{m,d,h}|$$
and the RMSE defined as
$$\textrm{RMSE}_m^{\textrm{test}} = \frac{1}{\#(\mathcal{D})} \sum_{d\in \mathcal{D}} \sqrt{\frac{1}{24} \sum_{h=0}^{23} (X_{m,d,h}-\what X_{m,d,h})^2}$$
where $\mathcal{D}$ is a set containing all 181 forecasted days, $\#(\cdot)$ is a function that returns the number of elements in a set, and $\what X_{m,d,h}$ represents the respective forecasted value. 

\begin{figure}[h!]
	\centering
	\subfigure[MAE]{\includegraphics[width = 0.8\textwidth,height= \textheight, keepaspectratio]{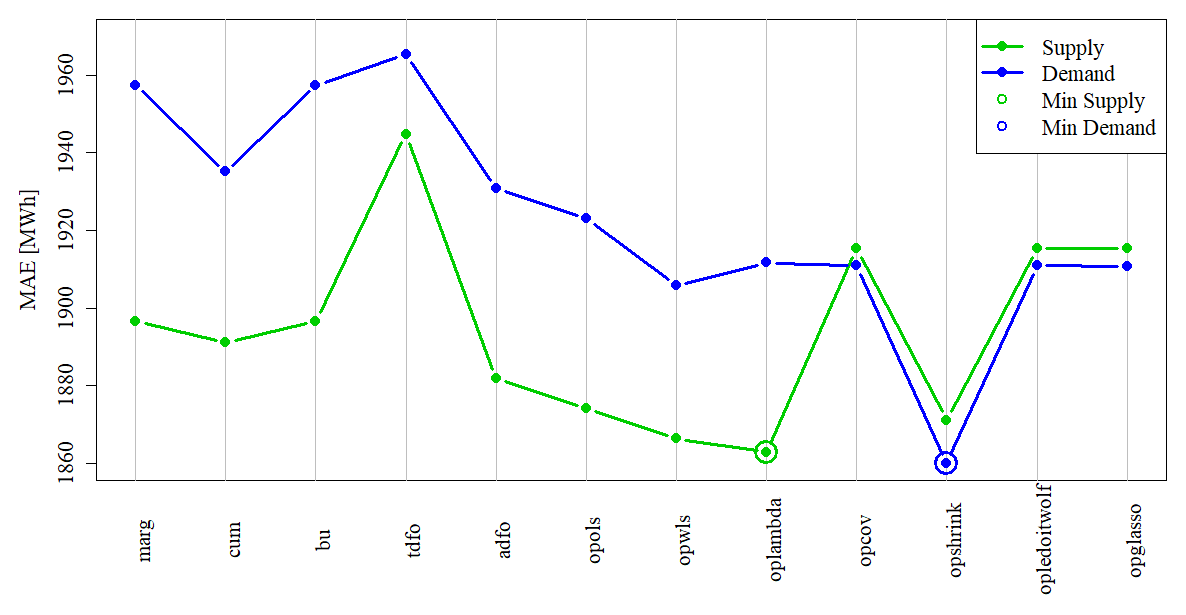}}
	\subfigure[RMSE]{\includegraphics[width = 0.8\textwidth,height= \textheight, keepaspectratio]{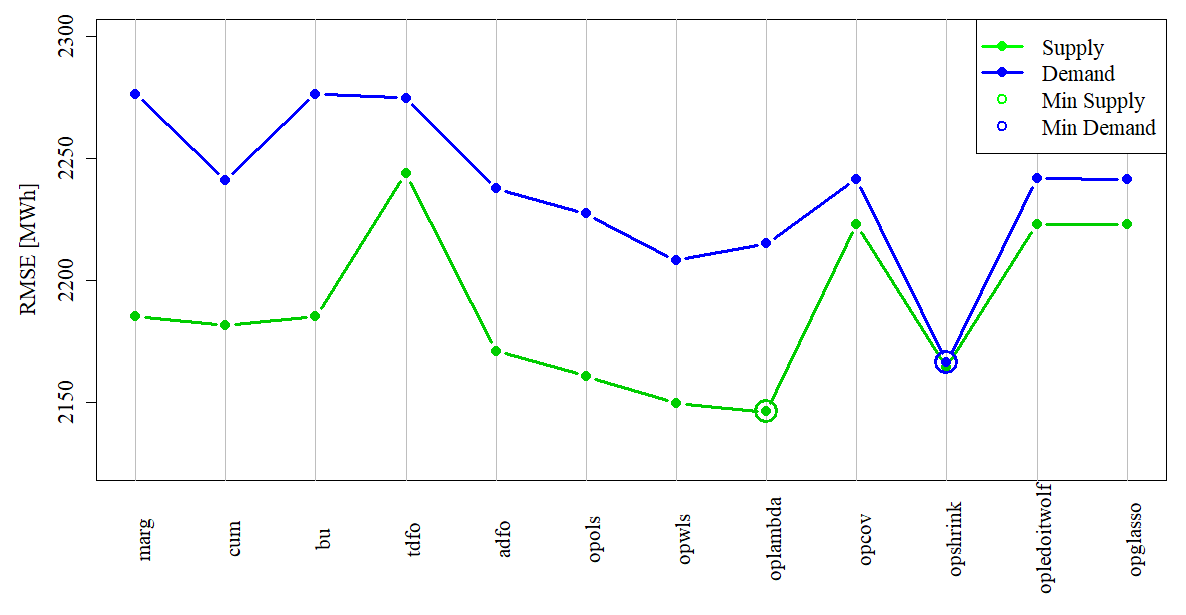}}
	\caption{Average MAE and RMSE over all considered reconciliation approaches for the Supply and Demand curves.}
	\label{fig:mae_rmse_fig}
\end{figure}

The averages over the MAEs and RMSEs for the overall price classes are shown in Figure \ref{fig:mae_rmse_fig}. The top-down and aggregated-down approaches using historical proportions yielded worse results than the simple bottom-up method. Therefore, we do not include these results but they can be inspected in the comprehensive Tables \ref{tab:resultsS} and \ref{tab:resultsD}. For the supply curve, the optimal lambda reconciliation approach obtained the smallest MAE on average, approximately 33 MWh lower than that with the marginal model. For the demand curve, the optimal shrinkage reconciliation approach returned the smallest MAE on average, approximatively 97 MWh lower than that of with the marginal model. Similar results were obtained in terms of RMSEs, with differences of approx. 39 MWh and 109 MWh for the supply and demand curves respectively.

\begin{table}[!htbp]
	\centering
	\resizebox{.99\textwidth}{!}{
\begin{tabular}{|r|rrrr|r|rrrr|r|c|}
	\hline
	& \multicolumn{5}{c|}{\bf Supply} & \multicolumn{5}{c|}{\bf Demand} & \bf Unit \\
	\textbf{Price class:} & -500.0 & -49.9 & 5.7 & 3000.0 & Mean & 3000.0 & 7.0 & -6.9 & -500.0 & Mean & \\ 
	\hline
	\hline
	marginal & 1591 & 1827 & 1941 & 2087 & 1897 & 2051 & 1951 & 19664 & 1958 & 1957 & MW \\ 
	cumulative & -3.478 & 0.084 & 0.015 & -0.584 & -0.290 & -2.540 & 0.907 & 1.307 & 2.689 & -1.141 & $\Delta$\% \\ 
	bu & 0.000 & 0.000 & 0.000 & 0.000 & 0.000 & 0.000 & 0.000 & 0.000 & 0.000 & 0.000 & $\Delta$\%  \\ 
	tdfo & 11.970 & 3.403 & 2.697 & -0.940 & 2.536 & -0.413 & 0.958 & 1.135 & 2.934 & 0.404 & $\Delta$\% \\ 
	adfo & 0.000 & -1.070 & -1.059 & -0.940 & -0.780 & 0.000 & 0.398 & 0.850 & 2.934 & -1.359 & $\Delta$\%\\ 
	opols & -0.472 & -1.496 & -1.667 & -1.163 & -1.183 & -1.349 & -0.409 & -0.054 & 2.655 & -1.756 & $\Delta$\% \\ 
	opwls & 3.021 & -1.684 & -2.120 & -2.607 & -1.597 & -2.860 & -2.495 & -2.379 & 1.053 & -2.633 & $\Delta$\% \\ 
	oplambda & -0.207 & -1.821 & -2.115 & -2.194 & -1.782 & -1.222 & -2.075 & -1.826 & 1.289 & -2.336 & $\Delta$\% \\ 
	opcov & 0.985 & 1.034 & 0.868 & -0.011 & 0.984 & -0.260 & -1.600 & -1.334 & -3.914 & -2.373 & $\Delta$\% \\ 
	opshrink & 0.289 & -1.668 & -1.899 & -1.785 & -1.346 & -4.102 & -4.443 & -4.235 & -4.168 & -4.979 & $\Delta$\% \\ 
	opledoitwolf & 0.986 & 1.035 & 0.869 & -0.010 & 0.984 & -0.261 & -1.598 & -1.331 & -3.906 & -2.371 & $\Delta$\% \\ 
	opglasso & 0.986 & 1.032 & 0.867 & -0.010 & 0.984 & -0.266 & -1.607 & -1.341 & -3.920 & -2.379 & $\Delta$\%  \\ 
	\hline
	\hline
	MAE Min & cumulative & oplambda & opwls & opwls & oplambda & opshrink & opshrink & opshrink & opshrink & opshrink & \\
	\hline
\end{tabular}}
	\caption{Absolute MAE and percentage differences relative to the base case (marginal) for selected classes and reconciliation methods. Here, the marginal case is the original implementation in the X-Model paper, which corresponds to the bottom-up (bu) case.}
	\label{tab:resultssum}
\end{table}

\PG{Table \ref{tab:resultssum} shows the absolute MAE and percentage differences for some selected classes and the mean over all classes. Again, we do not include the results for the top-down and aggregated-down approaches using historical proportions for the same reasons stated for Figure \ref{fig:mae_rmse_fig}. When using historical values to calculate the proportions, the top-down approach was superior to  aggregated-down. However, using forecasted values to calculate the proportions for the aggregated-down approach yielded better results than the bottom-up case on average. The top-down approach still yielded less accurate values than the bottom-up approach even when using forecasted values to calculate the proportions. Hence, aggregated-down was superior to bottom-up and top-down only when using forecasted values for the proportions.}

Tables \ref{tab:resultsS} and \ref{tab:resultsD} in the Appendix show the detailed MAEs and RMSEs over each price class. For the supply curve, no approach consistently yielded the lowest errors across all price classes. For example, the first three price classes had the lowest MAEs and RMSEs using the simple bottom-up approach. The cumulative and marginal models in the tables refer to simply forecasting all bottom-level values with the corresponding marginal or cumulative regressors and simply cumulatively summing them up, which corresponds to a bottom-up approach. It should be noted that in the results tables, the explicit bottom-up approach is equal to the cumulative approach per definition. The optimal weighted least squares (WLS) reconciliation approach yielded the lowest errors for the largest number of classes, followed by the shrinkage and lambda  approaches. For the demand curve, the results were consistent for all classes, where the lowest errors were achieved by the optimal shrinkage reconciliation approach (option 5 among the Section \hyperref[sec:opt]{3.3}).

\section{Conclusions}
\label{sec:conclusions}
\FZ{
In this study, we considered the hierarchical structure of aggregated curves and different representations.}
We presented several reconciliation methods comprising established bottom-up, top-down, and minimum-trace optimal reconciliation approaches in an aggregated curves setting \citep{wickramasuriya2019optimal}. In addition, we introduced a new \textit{aggregated-down} approach with comparable methodological complexity to the bottom-up and top-down approaches. \FZ{We provided a theoretical insight that under mild assumptions regarding the forecasting and reconciling method, the reconciling result is independent of the representation of the curve. 
These approaches were then applied in a simulation study, and for forecasting the supply and demand curves} of the German day-ahead electricity market. In the latter case, we showed that the reconciling approaches for aggregated curves can improve forecasting accuracy compared with the standard approaches.

The results showed that there is a single reconciliation method did not outperform the others every time. However, the Section \hyperref[sec:opt]{3.3} (optimal approaches), particularly the shrinkage, WLS, and lambda approaches, obtained the best results in most cases, where they obtained considerable improvements compared with the bottom-up base case. The reconciliation method that improves the forecast by the greates amount will probably be specific to the data.

 We suggest that it may be useful to consider multiple methods because even simple approaches such as Section \hyperref[sec:ad]{3.2} (aggregated-down) yielded improvements at certain points on the curve compared with the bottom-up approach, which is the current state-of-the-art method \citep{xmodel, ziel2018probabilistic, forecast2021haben}. 
 \PG{Based on our finding that the top-down approach did not improve the forecasts on average whereas the aggregated-down approach using forecasted values to calculate proportions obtained improvements, we conclude that the aggregated-down approach can be applied as a simple benchmark method that is superior to top-down. In addition, we conclude that it is important to have access to all base forecasts to calculate the proportions because they can lead to substantial improvements in the forecasts compared with only using historical values.}
 
 Our study could be extended  further by including more recent approaches, such as machine-learning-based \citep{spiliotis2021hierarchical} and conditional coherency reconciliation methods \citep{di2021forecast}. Averaging or using different reconciliation approaches for each point on the aggregated curve could also be considered, especially if coherency is not necessary.
 
\newpage
\section{Appendix}
\label{sec:appendix}

\FZ{
\subsection{Proof of Theorem 1}
Consider $\wtilde{\bsy} = \bsS \bsP \what{\bsy}$. With the definition of $\bsP$, $\bsS=\bsB_{[k]}\bsS_{[k]}\bsA_{[k]} \bsD_n^{-1}$, and the assumptions that $\what{\bsy} = \bsB_{[k]} \what{\bsy}_{[k]}$ and $\bsW^{-1}_{[k]} = \bsB_{[k]}'\bsW^{-1}\bsB_{[k]}$, it holds that: 
{\small 
\begin{align*}
	\wtilde{\bsy} &=	\bsS (\bsS'\bsW^{-1}\bsS)^{-1}\bsS' \bsW^{-1} \what{\bsy}\\
	&= \bsB_{[k]}\bsS_{[k]}\bsA_{[k]} \bsD_n^{-1} 
	\left((\bsB_{[k]}\bsS_{[k]}\bsA_{[k]} \bsD_n^{-1} )'\bsW^{-1} \bsB_{[k]}\bsS_{[k]}\bsA_{[k]} \bsD_n^{-1} \right)^{-1}(\bsB_{[k]}\bsS_{[k]}\bsA_{[k]} \bsD_n^{-1} )' \bsW^{-1}
	\what{\bsy} \\
	&= \bsB_{[k]}\bsS_{[k]}\bsA_{[k]} \bsD_n^{-1} 
	\left( \bsD_n \bsA_{[k]}^{-1} (\bsS_{[k]}'\bsB_{[k]}' \bsW^{-1} \bsB_{[k]}\bsS_{[k]})^{-1} (\bsA_{[k]}^{-1})' \bsD_n' \right) (\bsD_n^{-1})' \bsA_{[k]}' \bsS_{[k]}'\bsB_{[k]}' \bsW^{-1} 
	\what{\bsy} \\
	&= \bsB_{[k]}\bsS_{[k]} (\bsS_{[k]}'\bsB_{[k]}' \bsW^{-1} \bsB_{[k]}\bsS_{[k]})^{-1} \bsS_{[k]}'\bsB_{[k]}' \bsW^{-1} 
	 \bsB_{[k]} \what{\bsy}_{[k]} \\
	 &= \bsB_{[k]}\bsS_{[k]} (\bsS_{[k]}' \bsW^{-1}_{[k]} \bsS_{[k]})^{-1} \bsS_{[k]}'\bsW^{-1}_{[k]} \what{\bsy}_{[k]} = \bsB_{[k]} \wtilde{\bsy}_{[k]} .
\end{align*}
}
}

\subsection{Results tables}
\begin{table}[ht]
	\centering
	\scalebox{0.7}{
		\begin{tabular}{|r|rrr|rrr|rrr|}
			\hline
			\textbf{No. of bottom-level values ($n$)} & \multicolumn{3}{c|}{\textbf{4}} & \multicolumn{3}{c|}{\textbf{16}} & \multicolumn{3}{c|}{\textbf{64}}\\ \hline
			\textbf{No. of observations ($N$)} & \textbf{16} & \textbf{64} & \textbf{256} & \textbf{16} & \textbf{64} & \textbf{256} & \textbf{16} & \textbf{64} & \textbf{256} \\ 
			\hline
			base & 5.14 & 4.98 & 4.52 & 6.46 & 6.22 & 5.95 & 11.83 & 11.80 & 11.52 \\ 
			bu & 5.41 & 5.13 & 4.58 & 7.69 & 7.48 & 7.17 & 18.47 & 21.18 & 20.43 \\ 
			tdfo & 4.98 & 4.97 & 4.55 & 6.69 & 6.45 & 6.20 & 14.16 & 14.62 & 14.62 \\ 
			adfo & 5.08 & 4.97 & 4.52 & 6.39 & 6.23 & 5.97 & 11.80 & 11.82 & 11.54 \\ 
			opols & 5.09 & 4.97 & 4.51 & 6.42 & 6.20 & 5.93 & 11.81 & 11.79 & 11.50 \\ 
			opwls & 5.17 & 5.01 & 4.52 & 6.56 & 6.25 & 5.94 & 11.90 & 11.81 & 11.47 \\ 
			oplambda & 5.15 & 5.00 & 4.51 & 6.50 & 6.23 & 5.94 & 11.87 & 11.79 & 11.46 \\ 
			opshrink & 5.22 & 5.05 & 4.56 & 6.82 & 7.07 & 7.72 & 13.21 & 18.57 & 24.27 \\  
			\hline
		\end{tabular}
	}
	\caption{Simulation study RMSE results for the squared simulated values of the VAR(1) setup $\bsPhi = \sqrt{0.7}\bsI_n$ and $A=\bsI_n$ estimated by an AR(1) model, for selected reconciliation methods.}
	\label{tab:sim07abs}
\end{table}

\begin{table}[ht]
	\centering
	\scalebox{0.7}{
		\begin{tabular}{|r|rrr|rrr|rrr|}
			\hline
			\textbf{No. of bottom-level values ($n$)} & \multicolumn{3}{c|}{\textbf{4}} & \multicolumn{3}{c|}{\textbf{16}} & \multicolumn{3}{c|}{\textbf{64}}\\ \hline
			\textbf{No. of observations ($N$)} & \textbf{16} & \textbf{64} & \textbf{256} & \textbf{16} & \textbf{64} & \textbf{256} & \textbf{16} & \textbf{64} & \textbf{256} \\ 
			\hline
			base & 5.14 & 4.98 & 4.52 & 6.46 & 6.22 & 5.95 & 11.83 & 11.80 & 11.52 \\ 
			bu & 5.41 & 5.13 & 4.58 & 7.69 & 7.48 & 7.17 & 18.47 & 21.18 & 20.43 \\ 
			tdfo & 4.98 & 4.97 & 4.55 & 6.69 & 6.45 & 6.20 & 14.16 & 14.62 & 14.62 \\ 
			adfo & 5.08 & 4.97 & 4.52 & 6.39 & 6.23 & 5.97 & 11.80 & 11.82 & 11.54 \\ 
			opols & 5.09 & 4.97 & 4.51 & 6.42 & 6.20 & 5.93 & 11.81 & 11.79 & 11.50 \\ 
			opwls & 5.17 & 5.01 & 4.52 & 6.56 & 6.25 & 5.94 & 11.90 & 11.81 & 11.47 \\ 
			oplambda & 5.15 & 5.00 & 4.51 & 6.50 & 6.23 & 5.94 & 11.87 & 11.79 & 11.46 \\ 
			opshrink & 5.22 & 5.05 & 4.56 & 6.82 & 7.07 & 7.72 & 13.21 & 18.57 & 24.27 \\ 
			\hline
		\end{tabular}
	}
	\caption{Simulation study RMSE results for the VAR(1) setup $\bsPhi = 0.7\bsI_n$ and $A=\bsI_n$ for selected reconciliation methods, with outliers removed. Outliers were defined as simulations for which $\hat p_{fo,1}>50$.}
	\label{tab:sim07nooutliers}
\end{table}

\begin{table}[ht]
	\centering
	\scalebox{0.7}{
		\begin{tabular}{|r|rrr|rrr|rrr|}
			\hline
			\textbf{No. of bottom-level values ($n$)} & \multicolumn{3}{c|}{\textbf{4}} & \multicolumn{3}{c|}{\textbf{16}} & \multicolumn{3}{c|}{\textbf{64}}\\ \hline
			\textbf{No. of observations ($N$)} & \textbf{16} & \textbf{64} & \textbf{256} & \textbf{16} & \textbf{64} & \textbf{256} & \textbf{16} & \textbf{64} & \textbf{256} \\ 
			\hline
			base & 1.41 & 1.39 & 1.32 & 2.24 & 2.27 & 2.24 & 4.32 & 4.12 & 4.15 \\ 
			bu & 1.41 & 1.39 & 1.32 & 2.25 & 2.26 & 2.24 & 4.31 & 4.12 & 4.15 \\ 
			tdfo & 7.99 & 2.08 & 2.16 & 1700 & 16.26 & 13.03 & 163000 & 9040 & 15000 \\ 
			adfo & 1.42 & 1.39 & 1.32 & 2.26 & 2.27 & 2.24 & 4.33 & 4.13 & 4.15 \\ 
			opols & 1.40 & 1.39 & 1.32 & 2.23 & 2.27 & 2.24 & 4.31 & 4.12 & 4.15 \\ 
			opwls & 1.39 & 1.39 & 1.32 & 2.22 & 2.26 & 2.24 & 4.29 & 4.12 & 4.15 \\ 
			oplambda & 1.39 & 1.39 & 1.32 & 2.21 & 2.26 & 2.24 & 4.28 & 4.12 & 4.15 \\ 
			opshrink & 1.40 & 1.39 & 1.32 & 2.23 & 2.27 & 2.24 & 4.34 & 4.13 & 4.15 \\ 
			\hline
		\end{tabular}
	}
	\caption{Simulation study RMSE results for the VAR(1) setup $\bsPhi = 0.5\bsI_n$ and $A=\bsI_n$ for selected reconciliation methods}
	\label{tab:sim05}
\end{table}

\begin{table}[ht]
	\centering
	\scalebox{0.7}{
		\begin{tabular}{|r|rrr|rrr|rrr|}
			\hline
			\textbf{No. of bottom-level values ($n$)} & \multicolumn{3}{c|}{\textbf{4}} & \multicolumn{3}{c|}{\textbf{16}} & \multicolumn{3}{c|}{\textbf{64}}\\ \hline
			\textbf{No. of observations ($N$)} & \textbf{16} & \textbf{64} & \textbf{256} & \textbf{16} & \textbf{64} & \textbf{256} & \textbf{16} & \textbf{64} & \textbf{256} \\ 
			\hline
			base & 1.41 & 1.39 & 1.32 & 2.24 & 2.27 & 2.24 & 4.31 & 4.12 & 4.15 \\ 
			bu & 1.40 & 1.39 & 1.32 & 2.24 & 2.26 & 2.24 & 4.31 & 4.13 & 4.15 \\ 
			tdfo & 65.8 & 5.41 & 8.39 & 9670 & 590 & 889 & 9390000 & 4610 & 7220 \\ 
			adfo & 1.42 & 1.39 & 1.32 & 2.26 & 2.27 & 2.24 & 4.32 & 4.12 & 4.15 \\ 
			opols & 1.40 & 1.38 & 1.32 & 2.23 & 2.27 & 2.24 & 4.30 & 4.12 & 4.15 \\ 
			opwls & 1.40 & 1.38 & 1.32 & 2.22 & 2.26 & 2.24 & 4.28 & 4.11 & 4.15 \\ 
			oplambda & 1.39 & 1.38 & 1.32 & 2.22 & 2.26 & 2.24 & 4.27 & 4.11 & 4.15 \\ 
			opshrink & 1.40 & 1.38 & 1.32 & 2.23 & 2.26 & 2.24 & 4.33 & 4.13 & 4.15 \\
			\hline
		\end{tabular}
	}
	\caption{Simulation study RMSE results for the VAR(1) setup $\bsPhi = 0.2\bsI_n$ and $A=\bsI_n$ for selected reconciliation methods}
	\label{tab:sim02}
\end{table}

\begin{table}[ht]
	\centering
	\scalebox{0.7}{
		\begin{tabular}{|r|rrr|rrr|rrr|}
			\hline
			\textbf{No. of bottom-level values ($n$)} & \multicolumn{3}{c|}{\textbf{4}} & \multicolumn{3}{c|}{\textbf{16}} & \multicolumn{3}{c|}{\textbf{64}}\\ \hline
			\textbf{No. of observations ($N$)} & \textbf{16} & \textbf{64} & \textbf{256} & \textbf{16} & \textbf{64} & \textbf{256} & \textbf{16} & \textbf{64} & \textbf{256} \\ 
			\hline
			base & 1.41 & 1.39 & 1.32 & 2.26 & 2.26 & 2.24 & 4.35 & 4.14 & 4.15 \\ 
			bu & 1.40 & 1.39 & 1.32 & 2.27 & 2.26 & 2.23 & 4.38 & 4.13 & 4.15 \\ 
			tdfo & 14.9 & 2.71 & 1.55 & 31 & 20.7 & 574 & 16500 & 132 & 240 \\ 
			adfo & 1.43 & 1.39 & 1.33 & 2.29 & 2.26 & 2.24 & 4.38 & 4.14 & 4.15 \\ 
			opols & 1.39 & 1.39 & 1.32 & 2.24 & 2.26 & 2.24 & 4.34 & 4.14 & 4.15 \\ 
			opwls & 1.39 & 1.39 & 1.32 & 2.22 & 2.25 & 2.23 & 4.30 & 4.13 & 4.15 \\ 
			oplambda & 1.38 & 1.39 & 1.32 & 2.22 & 2.25 & 2.23 & 4.30 & 4.13 & 4.15 \\ 
			opshrink & 1.39 & 1.39 & 1.32 & 2.24 & 2.26 & 2.23 & 4.35 & 4.14 & 4.15 \\ 
			\hline
		\end{tabular}
	}
	\caption{Simulation study RMSE results for the VAR(1) setup $\bsPhi = 0.95\bsI_n$ and $A=\bsI_n$ for selected reconciliation methods}
	\label{tab:sim95}
\end{table}

\begin{table}[ht]
	\centering
	\scalebox{0.7}{
		\begin{tabular}{|r|rrr|rrr|rrr|}
			\hline
			\textbf{No. of bottom-level values ($n$)} & \multicolumn{3}{c|}{\textbf{4}} & \multicolumn{3}{c|}{\textbf{16}} & \multicolumn{3}{c|}{\textbf{64}}\\ \hline
			\textbf{No. of observations ($N$)} & \textbf{16} & \textbf{64} & \textbf{256} & \textbf{16} & \textbf{64} & \textbf{256} & \textbf{16} & \textbf{64} & \textbf{256} \\ 
			\hline
			base & 2.02 & 2.01 & 1.89 & 5.99 & 6.15 & 5.87 & 23.32 & 22.42 & 22.69 \\ 
			bu & 2.00 & 2.01 & 1.89 & 5.86 & 6.13 & 5.86 & 22.93 & 22.33 & 22.65 \\ 
			tdfo & 2.58 & 2.07 & 1.90 & 1440 & 6.19 & 5.97 & 30400000 & 23.2 & 23.4 \\ 
			adfo & 2.03 & 2.01 & 1.89 & 6.00 & 6.15 & 5.87 & 23.32 & 22.42 & 22.69 \\ 
			opols & 2.02 & 2.01 & 1.89 & 5.99 & 6.15 & 5.87 & 23.32 & 22.42 & 22.69 \\ 
			opwls & 2.01 & 2.01 & 1.89 & 5.93 & 6.14 & 5.87 & 23.12 & 22.37 & 22.67 \\ 
			oplambda & 2.01 & 2.01 & 1.89 & 5.97 & 6.15 & 5.87 & 23.30 & 22.41 & 22.69 \\ 
			opshrink & 2.01 & 2.01 & 1.89 & 5.92 & 6.15 & 5.87 & 23.08 & 22.32 & 22.68 \\
			\hline
		\end{tabular}
	}
	\caption{Simulation study RMSE results for the VAR(1) setup $\bsPhi = 0.7\bsI_n$ and $A= 0.3 \bsI_n + 0.7 \bsone \bsone'$ for selected reconciliation methods}
	\label{tab:simcorr}
\end{table}

\begin{sidewaystable}[ht]
	\centering
	\resizebox{.99\textwidth}{!}{
		\begin{tabular}{|rr|rrrrrrrrrrrrrrrrrrr|c|}
			\hline
			& & S-500.0 & S-171.5 & S-99.0 & S-76.0 & S-49.9 & S5.7 & S24.0 & S30.9 & S35.9 & S40.4 & S45.0 & S50.0 & S57.2 & S69.0 & S86.5 & S153.7 & S260.0 & S1600.0 & S3000.0 & Mean \\ 
			\hline
			\multirow{17}{*}{\rotatebox[origin=c]{90}{MAE}} & Marginal & 1591 & 1519 & \textbf{ 1503} & 1646 & 1827 & 1941 & 2034 & 2060 & 2010 & 1972 & 1949 & 1958 & 1956 & 1982 & 1980 & 1982 & 1981 & 2063 & 2087 & 1897 \\ 
			& Cumulative & \textbf{1535} & \textbf{1494} & 1519 & 1662 & 1829 & 1941 & 2031 & 2098 & 1987 & 1956 & 1956 & 1962 & 1935 & 1974 & 1967 & 1980 & 1980 & 2050 & 2075 & 1891 \\ 
			& bu & 1591 & 1519 & \textbf{1503} & 1646 & 1827 & 1941 & 2034 & 2060 & 2010 & 1971 & 1949 & 1957 & 1956 & 1982 & 1980 & 1982 & 1981 & 2063 & 2087 & 1897 \\ 
			& tdar & 3496 & 3191 & 2765 & 2354 & 2503 & 2680 & 2804 & 2820 & 2694 & 2532 & 2364 & 2277 & 2188 & 2064 & 2020 & 2031 & 2001 & 2047 & 2068 & 2468 \\ 
			& tdra & 3430 & 3142 & 2733 & 2358 & 2507 & 2671 & 2808 & 2829 & 2704 & 2546 & 2381 & 2289 & 2196 & 2072 & 2027 & 2037 & 2004 & 2048 & 2068 & 2466 \\ 
			& tdfo & 1781 & 1703 & 1666 & 1756 & 1889 & 1993 & 2087 & 2111 & 2045 & 2003 & 1972 & 1968 & 1954 & 1976 & 1973 & 1986 & 1975 & 2045 & 2068 & 1945 \\ 
			& adar & 1591 & 1498 & 1568 & 2171 & 2832 & 3189 & 3189 & 3117 & 2969 & 2882 & 2804 & 2848 & 2886 & 2966 & 2965 & 2975 & 3017 & 3141 & 3198 & 2727 \\ 
			& adra & 1591 & 1498 & 1568 & 2133 & 2769 & 3104 & 3122 & 3061 & 2910 & 2824 & 2753 & 2797 & 2842 & 2927 & 2934 & 2951 & 2997 & 3127 & 3186 & 2689 \\  
			& adfo  & 1591 & 1495 & 1518 & 1642 & 1808 & 1921 & 2009 & 2079 & 1972 & 1942 & 1941 & 1948 & 1922 & 1962 & 1957 & 1969 & 1970 & 2041 & 2068 & 1882 \\ 
			& opols & 1583 & 1506 & 1513 & 1635 & 1800 & 1909 & 2009 & 2043 & 1974 & 1941 & 1922 & 1928 & 1917 & 1951 & 1953 & 1967 & 1962 & 2037 & 2063 & 1874 \\ 
			& opwls & 1639 & 1553 & 1534 & 1643 & 1796 & \textbf{1900} & 1997 & 2021 & 1963 & \textbf{1928} & \textbf{1900} & \textbf{1901} & \textbf{1896} & \textbf{1931} & \textbf{1934} & \textbf{1947} &\textbf{1937} & \textbf{2009} & \textbf{2033} & 1866 \\ 
			& oplambda  & 1587 & 1511 & 1511 &  \textbf{1631} &  \textbf{1794} & 1900 & 1998 & 2024 & 1965 & 1931 & 1903 & 1906 & 1901 & 1936 & 1940 & 1954 & 1944 & 2017 & 2041 &  \textbf{1863} \\ 
			& opcov & 1606 & 1566 & 1557 & 1687 & 1846 & 1958 & 2051 & 2072 & 2010 & 1978 & 1956 & 1961 & 1963 & 2002 & 2007 & 2016 & 2002 & 2067 & 2087 & 1915 \\ 
			& opshrink & 1595 & 1535 & 1526 & 1644 & 1797 & 1904 & \textbf{1993} & \textbf{2017} & \textbf{1959} & 1928 & 1908 & 1910 & 1909 & 1952 & 1959 & 1972 & 1963 & 2029 & 2050 & 1871 \\  
			& opledoitwolf& 1606 & 1566 & 1557 & 1687 & 1846 & 1958 & 2051 & 2072 & 2010 & 1978 & 1956 & 1961 & 1963 & 2002 & 2007 & 2016 & 2002 & 2067 & 2087 & 1915 \\ 
			& opglasso & 1606 & 1566 & 1557 & 1687 & 1846 & 1958 & 2051 & 2072 & 2010 & 1978 & 1956 & 1961 & 1963 & 2002 & 2008 & 2016 & 2002 & 2067 & 2087 & 1915 \\ 
			\hline
			& MAE Minimum & Cumulative & Cumulative & Marginal & oplambda & oplambda & opwls & opshrink & opshrink & opshrink & opwls & opwls & opwls & opwls & opwls & opwls & opwls & opwls & opwls & opwls & oplambda \\ 
			\hline

			\multirow{17}{*}{\rotatebox[origin=c]{90}{RMSE}} & Marginal& 1846 & 1771 &  \textbf{1756} & 1911 & 2097 & 2228 & 2330 & 2361 & 2316 & 2273 & 2249 & 2254 & 2251 & 2282 & 2275 & 2279 & 2283 & 2367 & 2393 & 2185 \\ 
			& Cumulative & \textbf{1788} & \textbf{1746} & 1771 & 1926 & 2099 & 2230 & 2329 & 2397 & 2294 & 2256 & 2264 & 2262 & 2236 & 2276 & 2268 & 2284 & 2285 & 2355 & 2379 & 2181 \\  
			& bu & 1846 & 1771 & \textbf{1756} & 1911 & 2097 & 2228 & 2330 & 2361 & 2316 & 2273 & 2249 & 2254 & 2251 & 2282 & 2275 & 2279 & 2283 & 2367 & 2393 & 2185 \\ 
			& tdar& 3831 & 3522 & 3109 & 2680 & 2821 & 3021 & 3150 & 3161 & 3022 & 2855 & 2681 & 2590 & 2495 & 2363 & 2316 & 2331 & 2302 & 2350 & 2370 & 2788 \\ 
			& tdra & 3769 & 3477 & 3081 & 2685 & 2823 & 3012 & 3150 & 3167 & 3030 & 2866 & 2696 & 2601 & 2501 & 2368 & 2324 & 2338 & 2305 & 2351 & 2370 & 2785 \\
			& tdfo & 2051 & 1973 & 1945 & 2043 & 2179 & 2301 & 2397 & 2420 & 2357 & 2313 & 2283 & 2274 & 2257 & 2280 & 2274 & 2287 & 2278 & 2347 & 2370 & 2244 \\ 
			& adar & 1846 & 1749 & 1826 & 2447 & 3139 & 3519 & 3526 & 3460 & 3330 & 3244 & 3176 & 3229 & 3275 & 3370 & 3378 & 3394 & 3450 & 3589 & 3650 & 3084 \\ 
			& adra & 1846 & 1749 & 1826 & 2409 & 3073 & 3433 & 3455 & 3404 & 3271 & 3187 & 3123 & 3178 & 3230 & 3331 & 3349 & 3371 & 3431 & 3574 & 3637 & 3046 \\ 
			& adfo & 1846 & 1746 & 1771 & 1906 & 2077 & 2209 & 2307 & 2377 & 2275 & 2239 & 2246 & 2247 & 2223 & 2264 & 2256 & 2271 & 2274 & 2344 & 2370 & 2171 \\ 
			& opols & 1835 & 1754 & 1763 & 1898 & 2069 & 2197 & 2302 & 2340 & 2271 & 2235 & 2224 & 2224 & 2213 & 2249 & 2249 & 2265 & 2263 & 2338 & 2364 & 2161 \\ 
			& opwls & 1893 & 1804 & 1786 & 1909 & 2067 & 2189 & \textbf{2289} & \textbf{2314} & \textbf{2257} & \textbf{2218} & \textbf{2194} & \textbf{2192} & \textbf{2186} & \textbf{2222} & \textbf{2225} & \textbf{2238} & \textbf{2231} & \textbf{2303} & \textbf{2326} & 2150 \\ 
			& oplambda & 1840 & 1759 & 1760 & \textbf{1893} & \textbf{2062} & \textbf{2187} & 2289 & 2317 & 2259 & 2221 & 2199 & 2199 & 2193 & 2229 & 2232 & 2247 & 2240 & 2313 & 2337 & \textbf{2146} \\ 
			& opcov & 1881 & 1834 & 1829 & 1977 & 2143 & 2275 & 2375 & 2395 & 2331 & 2293 & 2274 & 2276 & 2277 & 2318 & 2320 & 2332 & 2319 & 2384 & 2405 & 2223 \\ 
			& opshrink & 1856 & 1792 & 1787 & 1919 & 2080 & 2207 & 2303 & 2326 & 2266 & 2229 & 2213 & 2212 & 2209 & 2253 & 2258 & 2270 & 2261 & 2329 & 2351 & 2164 \\ 
			& opledoitwolf & 1881 & 1834 & 1829 & 1977 & 2143 & 2275 & 2375 & 2395 & 2331 & 2293 & 2274 & 2276 & 2277 & 2318 & 2320 & 2332 & 2319 & 2384 & 2405 & 2223 \\ 
			& opglasso & 1881 & 1834 & 1829 & 1977 & 2143 & 2275 & 2375 & 2395 & 2331 & 2293 & 2274 & 2276 & 2277 & 2318 & 2320 & 2332 & 2319 & 2384 & 2405 & 2223 \\ 
			\hline
			& RMSE Minimum & Cumulative & Cumulative & Marginal & oplambda & oplambda & oplambda & opwls & opwls & opwls & opwls & opwls & opwls & opwls & opwls & opwls & opwls & opwls & opwls & opwls & oplambda \\ 
			\hline
	\end{tabular}}
	\caption{MAE and RMSE for the Supply curve and all considered reconciliation methods}
	\label{tab:resultsS}
\end{sidewaystable}

\begin{sidewaystable}[ht]
	\centering
	\resizebox{.99\textwidth}{!}{
		\begin{tabular}{|rr|rrrrrrrrrrrrrrrrrrrrr|c|}
			\hline
			& & D3000.0 & D1001.1 & D250.1 & D155.1 & D76.0 & D50.1 & D42.1 & D35.5 & D30.1 & D25.1 & D19.1 & D11.1 & D7.0 & D-6.9 & D-58.9 & D-76.9 & D-90.0 & D-146.0 & D-180.0 & D-388.8 & D-500.0 & Mean \\ 
			\hline
			\multirow{17}{*}{\rotatebox[origin=c]{90}{MAE}} & Marginal & 2051 & 2042 & 2005 & 2003 & 1962 & 1950 & 1964 & 1946 & 1961 & 1969 & 1958 & 1952 & 1951 & 1966 & 2005 & 1906 & 1818 & 1883 & 1913 & 1942 & 1957 & 1957 \\ 
			& Cumulative & 1999 & 1998 & 1945 & 1932 & 1893 & 1869 & 1883 & 1866 & 1874 & 1937 & 1950 & 1946 & 1968 & 1992 & 2039 & 1893 & 1803 & 1894 & 1961 & 1984 & 2010 & 1935 \\ 
			& bu & 2051 & 2042 & 2005 & 2003 & 1962 & 1950 & 1964 & 1946 & 1961 & 1969 & 1958 & 1952 & 1951 & 1966 & 2005 & 1906 & 1818 & 1883 & 1913 & 1942 & 1957 & 1957 \\ 
			& tdar & 2781 & 2790 & 2827 & 2801 & 2616 & 2481 & 2343 & 2210 & 2116 & 2161 & 2070 & 1940 & 1911 & 1934 & 1944 & 1895 & 1816 & 1899 & 1983 & 2021 & 2015 & 2217 \\ 
			& tdra & 2767 & 2775 & 2812 & 2784 & 2602 & 2473 & 2340 & 2209 & 2120 & 2170 & 2083 & 1945 & 1910 & 1932 & 1941 & 1892 & 1816 & 1901 & 1985 & 2022 & 2015 & 2214 \\ 
			& tdfo & 2042 & 2032 & 1976 & 1973 & 1944 & 1929 & 1945 & 1925 & 1939 & 1960 & 1961 & 1966 & 1969 & 1989 & 2040 & 1929 & 1845 & 1930 & 1967 & 1994 & 2015 & 1965 \\ 
			& adar & 2051 & 2043 & 2028 & 1991 & 1932 & 1965 & 2011 & 2104 & 2244 & 2344 & 2321 & 2517 & 2829 & 2983 & 2940 & 2953 & 2779 & 2806 & 2911 & 3013 & 3042 & 2467 \\ 
			& adra & 2051 & 2044 & 2032 & 1995 & 1934 & 1963 & 2001 & 2085 & 2215 & 2309 & 2280 & 2460 & 2765 & 2922 & 2873 & 2879 & 2713 & 2744 & 2847 & 2942 & 2968 & 2430 \\ 
			& adfo & 2051 & 1996 & 1944 & 1928 & 1889 & 1863 & 1869 & 1851 & 1854 & 1915 & 1926 & 1930 & 1958 & 1983 & 2030 & 1886 & 1804 & 1899 & 1967 & 1987 & 2015 & 1931 \\ 
			& opols & 2023 & 1996 & 1940 & 1925 & 1879 & 1854 & 1862 & 1845 & 1862 & 1901 & 1911 & 1923 & 1943 & 1965 & 2018 & 1890 & 1803 & 1898 & 1951 & 1985 & 2009 & 1923 \\ 
			& opwls & 1992 & 1984 & 1932 & 1917 & 1874 & 1855 & 1868 & 1851 & 1867 & 1894 & 1893 & 1897 & 1902 & 1919 & 1979 & 1869 & 1781 & 1887 & 1927 & 1956 & 1978 & 1906 \\ 
			& oplambda & 2026 & 1999 & 1942 & 1926 & 1878 & 1855 & 1865 & 1847 & 1863 & 1891 & 1892 & 1900 & 1910 & 1930 & 1991 & 1880 & 1794 & 1886 & 1929 & 1960 & 1983 & 1912 \\ 
			& opcov & 2045 & 2031 & 1975 & 1932 & 1900 & 1883 & 1890 & 1874 & 1895 & 1925 & 1919 & 1918 & 1919 & 1940 & 1994 & 1866 & 1788 & 1835 & 1851 & 1869 & 1881 & 1911 \\ 
			& opshrink &  \textbf{1967} &  \textbf{1957} &  \textbf{1904} &  \textbf{1879} &  \textbf{1844} &  \textbf{1823} &  \textbf{1831} &  \textbf{1818} &  \textbf{1833} &  \textbf{1863} &  \textbf{1860} &  \textbf{1860} &  \textbf{1864} &  \textbf{1883} &  \textbf{1940} &  \textbf{1823} &  \textbf{1739} &  \textbf{1803} &  \textbf{1834} &  \textbf{1860} &  \textbf{1876} &  \textbf{1860} \\ 
			& opledoitwolf & 2045 & 2031 & 1975 & 1932 & 1900 & 1883 & 1890 & 1874 & 1895 & 1925 & 1919 & 1919 & 1920 & 1940 & 1994 & 1866 & 1788 & 1835 & 1851 & 1869 & 1881 & 1911 \\ 
			& opglasso & 2045 & 2031 & 1975 & 1932 & 1900 & 1883 & 1890 & 1874 & 1894 & 1924 & 1918 & 1918 & 1919 & 1940 & 1993 & 1866 & 1788 & 1835 & 1851 & 1869 & 1881 & 1911 \\ 
			\hline
			& MAE Minimum	& opshrink & opshrink & opshrink & opshrink & opshrink & opshrink & opshrink & opshrink & opshrink & opshrink & opshrink & opshrink & opshrink & opshrink & opshrink & opshrink & opshrink & opshrink & opshrink & opshrink & opshrink & opshrink \\ 
			\hline

			\multirow{17}{*}{\rotatebox[origin=c]{90}{RMSE}} & Marginal & 2377 & 2367 & 2337 & 2329 & 2284 & 2273 & 2284 & 2267 & 2276 & 2284 & 2276 & 2273 & 2272 & 2291 & 2328 & 2224 & 2133 & 2194 & 2221 & 2247 & 2263 & 2276 \\ 
			& Cumulative & 2324 & 2320 & 2269 & 2257 & 2210 & 2190 & 2200 & 2174 & 2179 & 2241 & 2249 & 2244 & 2268 & 2298 & 2343 & 2188 & 2100 & 2183 & 2251 & 2272 & 2301 & 2241 \\ 
			& bu & 2377 & 2367 & 2337 & 2329 & 2284 & 2273 & 2284 & 2267 & 2276 & 2284 & 2276 & 2273 & 2272 & 2291 & 2328 & 2224 & 2133 & 2194 & 2221 & 2247 & 2263 & 2276 \\ 
			& tdar & 3129 & 3140 & 3173 & 3142 & 2948 & 2802 & 2673 & 2541 & 2443 & 2488 & 2393 & 2263 & 2234 & 2259 & 2262 & 2207 & 2120 & 2195 & 2278 & 2310 & 2305 & 2538 \\ 
			& tdra & 3117 & 3127 & 3161 & 3127 & 2935 & 2795 & 2671 & 2540 & 2447 & 2498 & 2406 & 2269 & 2233 & 2258 & 2260 & 2205 & 2120 & 2197 & 2279 & 2311 & 2305 & 2536 \\ 
			& tdfo & 2359 & 2350 & 2299 & 2291 & 2258 & 2246 & 2264 & 2244 & 2253 & 2276 & 2274 & 2277 & 2281 & 2302 & 2348 & 2230 & 2147 & 2223 & 2259 & 2284 & 2305 & 2275 \\ 
			& adar & 2377 & 2370 & 2356 & 2311 & 2250 & 2298 & 2350 & 2448 & 2586 & 2685 & 2660 & 2847 & 3149 & 3305 & 3264 & 3270 & 3099 & 3109 & 3215 & 3322 & 3351 & 2791 \\ 
			& adra & 2377 & 2371 & 2359 & 2315 & 2251 & 2294 & 2340 & 2429 & 2558 & 2650 & 2620 & 2791 & 3088 & 3247 & 3199 & 3199 & 3033 & 3049 & 3154 & 3253 & 3278 & 2755 \\ 
			& adfo & 2377 & 2317 & 2269 & 2254 & 2207 & 2186 & 2189 & 2160 & 2160 & 2220 & 2227 & 2230 & 2258 & 2289 & 2333 & 2182 & 2104 & 2188 & 2257 & 2275 & 2305 & 2237 \\ 
			& opols & 2344 & 2313 & 2263 & 2246 & 2195 & 2173 & 2178 & 2155 & 2166 & 2203 & 2209 & 2222 & 2241 & 2269 & 2319 & 2183 & 2098 & 2186 & 2240 & 2272 & 2298 & 2227 \\ 
			& opwls & 2307 & 2297 & 2252 & 2235 & 2186 & 2167 & 2178 & 2160 & 2172 & 2196 & 2192 & 2196 & 2201 & 2224 & 2280 & 2161 & 2075 & 2174 & 2213 & 2241 & 2262 & 2208 \\ 
			& oplambda & 2347 & 2316 & 2264 & 2245 & 2192 & 2171 & 2177 & 2157 & 2168 & 2193 & 2191 & 2199 & 2209 & 2235 & 2291 & 2172 & 2087 & 2174 & 2215 & 2245 & 2267 & 2215 \\ 
			& opcov & 2401 & 2385 & 2329 & 2277 & 2235 & 2217 & 2223 & 2208 & 2226 & 2255 & 2245 & 2243 & 2243 & 2268 & 2319 & 2186 & 2103 & 2153 & 2170 & 2187 & 2200 & 2242 \\ 
			& opshrink & \textbf{2286} & \textbf{2274} & \textbf{2224} & \textbf{2197} & \textbf{2155} & \textbf{2136} & \textbf{2144} & \textbf{2130} & \textbf{2144} & \textbf{2171} & \textbf{2166} & \textbf{2167} & \textbf{2169} & \textbf{2192} & \textbf{2245} & \textbf{2121} & \textbf{2039} & \textbf{2098} & \textbf{2125} & \textbf{2148} & \textbf{2164} & \textbf{2166} \\ 
			& opledoitwolf & 2401 & 2385 & 2329 & 2277 & 2235 & 2217 & 2223 & 2208 & 2227 & 2255 & 2245 & 2243 & 2243 & 2268 & 2319 & 2186 & 2103 & 2153 & 2170 & 2187 & 2200 & 2242 \\ 
			& opglasso & 2401 & 2385 & 2329 & 2276 & 2235 & 2217 & 2223 & 2208 & 2226 & 2254 & 2245 & 2243 & 2242 & 2268 & 2319 & 2186 & 2103 & 2153 & 2170 & 2187 & 2199 & 2241 \\ 
			\hline
			& RMSE Minimum & opshrink & opshrink & opshrink & opshrink & opshrink & opshrink & opshrink & opshrink & opshrink & opshrink & opshrink & opshrink & opshrink & opshrink & opshrink & opshrink & opshrink & opshrink & opshrink & opshrink & opshrink & opshrink \\
			\hline
	\end{tabular}}
	\caption{MAE and RMSE for the Demand curve and all considered reconciliation methods}
	\label{tab:resultsD}
\end{sidewaystable}

\clearpage
\vspace{-5mm} 
 \bibliographystyle{unsrtnat1}

\bibliography{bibliography}

\end{document}